\documentclass[traditabstract]{aa}
\usepackage{graphics}
\usepackage{graphicx}
\usepackage{multicol}
\usepackage{longtable}
\usepackage{txfonts}

\usepackage{natbib,twoopt}
\usepackage{amsmath} 
\usepackage[breaklinks=true]{hyperref} 
\bibpunct{(}{)}{;}{a}{}{,} 
\makeatletter
\newcommandtwoopt{\citeads}[3][][]{\href{http://adsabs.harvard.edu/abs/#3}%
{\def\hyper@linkstart##1##2{}%
\let\hyper@linkend\@empty\citealp[#1][#2]{#3}}}
\newcommandtwoopt{\citepads}[3][][]{\href{http://adsabs.harvard.edu/abs/#3}%
{\def\hyper@linkstart##1##2{}%
\let\hyper@linkend\@empty\citep[#1][#2]{#3}}}
\newcommandtwoopt{\citetads}[3][][]{\href{http://adsabs.harvard.edu/abs/#3}%
{\def\hyper@linkstart##1##2{}%
\let\hyper@linkend\@empty\citet[#1][#2]{#3}}}
\newcommandtwoopt{\citeyearads}[3][][]%
{\href{http://adsabs.harvard.edu/abs/#3}
{\def\hyper@linkstart##1##2{}%
\let\hyper@linkend\@empty\citeyear[#1][#2]{#3}}}
\makeatother

\usepackage{color}
\definecolor{mygreen}{RGB}{0,128,0}

\hypersetup{colorlinks=true,linkcolor=blue,citecolor=blue,urlcolor=blue}

\def\parallax{15.61}
\def\parallaxerr{0.99}
\def\deltalphaB{-30.0}
\def\deltdeltaB{-6.0}

\def\rhoabsB{30.6}
\def\rhoBerr{2.5}
\def\RB{1.96}
\def\RBerr{0.16}
\def\massBjup{12}
\def\massBjuperr{16}
\def\masssini{0.653}
\def\masssinierrstat{0.011}
\def\masssinierrparallax{0.041}
\def\mass{0.659}
\def\masserrstat{0.011}
\def\masserrparallax{0.041}
\def\masserrtot{0.043}
\def\masserrpercent{6.6\%}
\def\radius{123}
\def\radiuserr{14}
\def\luminosity{2000}
\def\luminosityerr{700}
\def\efftemp{3500}
\def\efftemperr{250}
\def\period{138.3}
\def\perioderr{1.7}
\def\logg{0.078}
\def\loggerr{0.027}
\def\subkepA{40.7}
\def\subkepAerr{5.5}
\def\subkealpha{-0.853}
\def\subkepalphaerr{0.059}

\begin{document}

\title{ALMA observations of the nearby AGB star L$_2$\,Puppis}
\subtitle{I. Mass of the central star and detection of a candidate planet}
\titlerunning{Mass and candidate planet of L$_2$\,Pup from ALMA}
\authorrunning{P. Kervella et al.}
%
\author{
P.~Kervella\inst{1,2}
\and
W.~Homan\inst{3}
\and
A.~M.~S.~Richards\inst{4}
\and
L.~Decin\inst{3}
\and
I.~McDonald\inst{4}
\and
M.~Montarg\`es\inst{5}
\and
K.~Ohnaka\inst{6}
}
\institute{
Unidad Mixta Internacional Franco-Chilena de Astronom\'{i}a (CNRS UMI 3386), Departamento de Astronom\'{i}a, Universidad de Chile, Camino El Observatorio 1515, Las Condes, Santiago, Chile, \email{pkervell@das.uchile.cl}.
\and
LESIA (UMR 8109), Observatoire de Paris, PSL Research University, CNRS, UPMC, Univ. Paris-Diderot, 5 Place Jules Janssen, 92195 Meudon, France, \email{pierre.kervella@obspm.fr}.
\and
Institute of Astronomy, KU Leuven, Celestijnenlaan 200D B2401, 3001 Leuven, Belgium
\and
JBCA, Department Physics and Astronomy, University of Manchester, Manchester M13 9PL, United Kingdom
\and
Institut de Radioastronomie Millim\'etrique, 300 rue de la Piscine, 38406, Saint Martin d'H\`eres, France
\and
Universidad Cat\'olica del Norte, Instituto de Astronom\'ia, Avenida Angamos 0610, Antofagasta, Chile
}
\date{Received ; Accepted}
\abstract
   {
   Six billion years from now, while evolving on the asymptotic giant branch (AGB), the Sun will metamorphose from a red giant into a beautiful planetary nebula.
   This spectacular evolution will impact the solar system planets, but observational confirmations of the predictions of evolution models are still elusive as no planet orbiting an AGB star has yet been discovered.
   The nearby AGB red giant L$_2$\,Puppis ($d=64$\,pc) is surrounded by an almost edge-on circumstellar dust disk.
   We report new observations with ALMA at very high angular resolution ($18 \times 15$\,mas) in band 7 ($\nu \approx 350$\,GHz) that allow us to resolve the velocity profile of the molecular disk.
   We establish that the gas velocity profile is Keplerian within the central cavity of the dust disk,  allowing us to derive the mass of the central star L$_2$\,Pup~A, $m_A = \mass \pm \masserrstat \pm \masserrparallax\,M_\odot$ ($\pm \masserrpercent$).
   From evolutionary models, we determine that L$_2$\,Pup A had a near-solar main-sequence mass, and is therefore a close analog of the future Sun in 5 to 6\,Gyr.
   The continuum map reveals a secondary source (B) at a radius of 2\,AU contributing $f_B/f_A = 1.3 \pm 0.1\%$ of the flux of the AGB star.
   L$_2$\,Pup~B is also detected in CO emission lines at a radial velocity of $v_B = 12.2 \pm 1.0$\,km\ s$^{-1}$.
   The close coincidence of the center of rotation of the gaseous disk with the position of the continuum emission from the AGB star allows us to constrain the mass of the companion to $m_B = \massBjup \pm \massBjuperr\,M_\mathrm{Jup}$.
   L$_2$\,Pup~B is most likely a planet or low-mass brown dwarf with an orbital period of about five years.
   Its continuum brightness and molecular emission suggest that it may be surrounded by an extended molecular atmosphere or an accretion disk.
  L$_2$\,Pup therefore emerges as a promising vantage point on the distant future of our solar system.
   }
\keywords{Stars: individual: HD 56096; Stars: AGB and post-AGB; Stars: circumstellar matter; Techniques: high angular resolution; Planetary systems; Planets and satellites: detection}

\maketitle



\section{Introduction}

Planets are ubiquitous at all stages of stellar evolution: from young stellar objects \citepads{David:2016aa}, main-sequence stars \citepads{1995Natur.378..355M,2010Sci...327..977B}, red giants \citepads{2016arXiv160605818G}, up to white dwarfs and neutron stars \citepads{1994Sci...264..538W, 2003Sci...301..193S, 2006Natur.442..543M}.
The asymptotic giant branch (hereafter AGB) designates the brief phase of the evolution of low- and intermediate-mass stars during which they metamorphose from red giants into compact stellar remnants, experiencing intense mass loss and extreme changes in their brightness and temperature.
This directly affects their planetary systems, in the most dramatic way since their formation. Their fate during the final stellar evolution stages has been the subject of several recent works (\citeads{2016MNRAS.463.2958V, 2016MNRAS.458..832S, 2012MNRAS.421.2969V, 2008MNRAS.386..155S, 2007ApJ...661.1192V}; see also the review by \citeads{2016RSOS....350571V}).
However, as AGB star planets are embedded in complex circumstellar envelopes and are vastly outshone by their parent star, the observation of this critical phase presents considerable and yet unsolved challenges.
As a result, there currently exists only indirect evidence of planets orbiting AGB stars \citepads{2009A&A...498..801W}.

At a distance of $64 \pm 4$\,pc ($\pi = \parallax \pm \parallaxerr$\,mas, \citeads{2007A&A...474..653V}) \object{L$_2$\,Puppis} (\object{HD 56096}, \object{HIP 34922}, \object{HR 2748}, \object{2MASS J07133229-4438233}) is the second nearest AGB star behind R\,Doradus ($\pi = 18.31 \pm 0.99$\,mas), and it is $\approx 30\%$ closer than \object{Mira}.
L$_2$\,Pup is a semi-regular pulsating variable ($P \approx 140$\,days, \citeads{1985IBVS.2681....1K, 2005MNRAS.361.1375B}).
An asymmetric resolved environment around L$_2$\,Pup was first identified by \citetads{2004MNRAS.350..365I} using aperture masking in the optical and near-infrared.
\citetads{2014A&A...564A..88K} observed L$_2$\,Pup in 2013 using the VLT/NACO adaptive optics (AO) between 1.0 and 4.0\,$\mu$m, detecting an edge-on circumstellar dust disk.
As the scattering of the stellar light by the dust is more efficient at shorter wavelengths, the central source appeared obscured by a dark band up to $\lambda \approx 1.2\,\mu$m ($J$ band).
At longer wavelengths, the scattering becomes less efficient and the transparent dust lets the thermal emission from the hot ($\approx 1000$\,K) inner rim of the dust disk pass. In the $L$ band ($\lambda = 4\,\mu$m), the thermal emission from a loop extending to $\approx 10$\,AU was also observed.
From NACO aperture masking and long-baseline interferometry, \citetads{2015A&A...576A..46L, 2015A&A...581C...2L} and \citetads{2015A&A...581A.127O} confirmed the overall geometry and extension of the disk.
\citetads{2015A&A...578A..77K} identified the polarimetric signature of the circumstellar disk and detected bipolar ``hourglass'' cones in L$_2$\,Pup's envelope using the VLT/SPHERE AO imaging polarimeter.
They also discovered streamers in the bipolar cones and two thin, tightly collimated plumes.
These structures make L$_2$\,Pup a promising candidate for hosting a low-mass companion, as circumstellar disks and bipolar cones are classically predicted by hydrodynamical models of binary objects \citepads{2006MNRAS.370.2004N}.
\citetads{2016MNRAS.460.4182C} presented the results of a 3D hydrodynamical simulation of L$_2$\,Pup as a binary object that reproduces the spectral energy distribution and morphology of the disk.
%

Knowing the mass of the central star in L$_2$\,Pup is essential to constrain its age and evolutionary state.
Indirect estimates vary considerably in the literature: from $0.5 \pm 0.2\,M_\odot$ by \citetads{2014A&A...561A..47O}, $0.7\,M_\odot$ by \citetads{2015A&A...576A..46L}, and $1.7\,M_\odot$ by \citetads{1998NewA....3..137D} to $2\,M_\odot$ by \citetads{2014A&A...564A..88K}.
With the goal of precisely determining the mass of the AGB star, we obtained very high angular resolution observations in ALMA's band 7 ($330-360$\,GHz), covering several molecular lines and the continuum (Sect.~\ref{observations}).
The simultaneous high spatial and spectral resolution provided by ALMA allows us to map the kinematics of its molecular envelope and derive its mass (Sect.~\ref{analysis}).
We also detect a secondary source located at a projected separation of 2\,AU from the primary.
We discuss the evolutionary state of the AGB star and the nature of the companion source in Sect.~\ref{discussion}.

\section{Observations and data reduction}\label{observations}

\begin{table*}
        \caption{Selected ALMA spectral windows (spw). The sky frequencies assume a source radial velocity of $v = 33.0$\,km\,s$^{-1}$.}
        \centering          
        \label{l2pup-spw}
        \begin{tabular}{clllcrcc}
	\hline\hline
        \noalign{\smallskip}
Baseband & Objective & Rest $\nu$ & Sky $\nu$ & Number of & Channel width & spw width & Velocity res. \\
& & (GHz) & (GHz) &  channels & (MHz) & (MHz) & (km\,s$ ^{-1}$) \\
        \noalign{\smallskip}
        \hline    
        \noalign{\smallskip}
1 & $^{12}$CO & 345.79599 & 345.75793 &  3840 & 0.122073 & 469 & 0.106 \\
        \hline    
        \noalign{\smallskip}
2 & $^{29}$SiO & 342.98085 & 342.94309 &  960 & 0.488292 & 469 & 0.427 \\
 & HC$^{15}$N & 344.20011 & 344.16222 & 480 & 0.488292 & 234 & 0.426 \\
        \hline    
        \noalign{\smallskip}
3 & $^{13}$CO & 330.58797 & 330.55158 & 1920 & 0.244146 & 469 & 0.222 \\
        \hline    
        \noalign{\smallskip}
4 & Continuum & 331.60000 & 331.56350 & 128 & 15.625350 & 2000 & 14.136 \\
        \hline
      \end{tabular}
\end{table*}

L$_2$\,Pup was observed on 5 November 2015 at UT08:39:25 (epoch 2015.8448, $\mathrm{MJD} = 57331.361$) by ALMA for project code 2015.1.00141.S.
Forty-five antennas were present for most of the observations, providing baselines from 0.09 to 16 km and resulting in an angular resolution better than 15\,mas.
The coverage of the $(u,v)$ plane inside 350\,m is sparse, giving a maximum angular scale of about 200\,mas for reliable imaging, which is a good match to the extension of the disk surrounding L$_2$\,Pup.
The total field of view to the half-power primary beam is approximately $15\arcsec$.
The observations used spectral windows (spw) placed as described in Table~\ref{l2pup-spw}.
Two identical executions of the high-resolution band 7 science goal were performed in succession. The observing frequency of each spw, corrected for the Earth's motion with respect to the Local Standard of Rest (LSR) and the VLSR of L$_2$ Pup (33.0\,km\,s$ ^{-1}$) was calculated for the start of each execution and then held fixed.
The target was observed for a total of 84 min, alternating with the phase reference with a cadence of 90\,s / 18\,s; the total cycle time was 2-3 min allowing for system temperature and water vapor radiometry measurements and occasional scans on the check source. The precipitable water vapor was very low during observations, at $0.40-0.45$\,mm.
Standard human-steered ALMA data reduction scripts were used \citepads{2014SPIE.9149E..0ZS}.
The main stages are to apply the instrumental calibration, to flag edge channels and other bad data, and to use the astrophysical calibrators to derive the bandpass corrections, flux scale and time-dependent phase and amplitude corrections.
These were applied to L$_2$\,Pup and to the check source.

The compact QSOs J0538-4405, J0701-4634 and J0726-4728 were used to establish the flux scale and derive bandpass corrections as the phase reference source and as the check source, respectively.
L$_2$\,Pup and J0726-4728 have separations of $2.85^\circ$ and $4.33^\circ$ from the phase-reference J0701-4634.
The flux density of J0538-4405 was taken as 1.04488 Jy at 338.994438 GHz, spectral index $-0.596$, based on fortnightly ALMA flux monitoring derived from planetary standards, with 5\% accuracy at band 7 frequency.
The derived flux densities of J0701-4634 and J0726-4728 were $0.396 \pm 0.002$\,Jy and $0.145 \pm 0.007$\,Jy.
Assuming that the flux scale transfer to L$_2$\,Pup is of similar accuracy to that of the check source, the overall flux scale accuracy is 7\%.

After applying the phase-reference and other calibrations to L$_2$\,Pup, the corrected target data were split out and each spectral window adjusted to fixed velocity with respect to the LSR. Obvious spectral lines were identified in the visibility data, leaving 2.5\,GHz of line-free continuum.
A copy of the data with all channels averaged to the coarsest resolution was made to speed up continuum imaging.
The continuum image (made with natural weighting) has a synthesized beam size $17.7 \times 14.5$\,mas at position angle (PA) 73$^\circ$.

The position of the continuum peak was located at $\alpha = 07$:13:32.47687, $\delta = -44$:38:17.8443 with an absolute position uncertainty of $\pm 7$\,mas.
The clean components of this image were used as a model for phase self-calibration, and iterative cycles of phase and amplitude self-calibration were performed.
Multi-frequency synthesis was used with a linear position-dependent spectral index as a free parameter; although the spectral index is not reliable except for the brightest emission over the relatively narrow, unevenly sampled bandwidth, this improves the image fidelity.
A 2.5\,mas pixel size and a field of view of $2.56\arcsec$ were used for all images unless otherwise stated.

\section{Analysis \label{analysis}}

\subsection{Continuum emission\label{continuum}}

\begin{figure*}[]
        \centering
        \includegraphics[width=8cm]{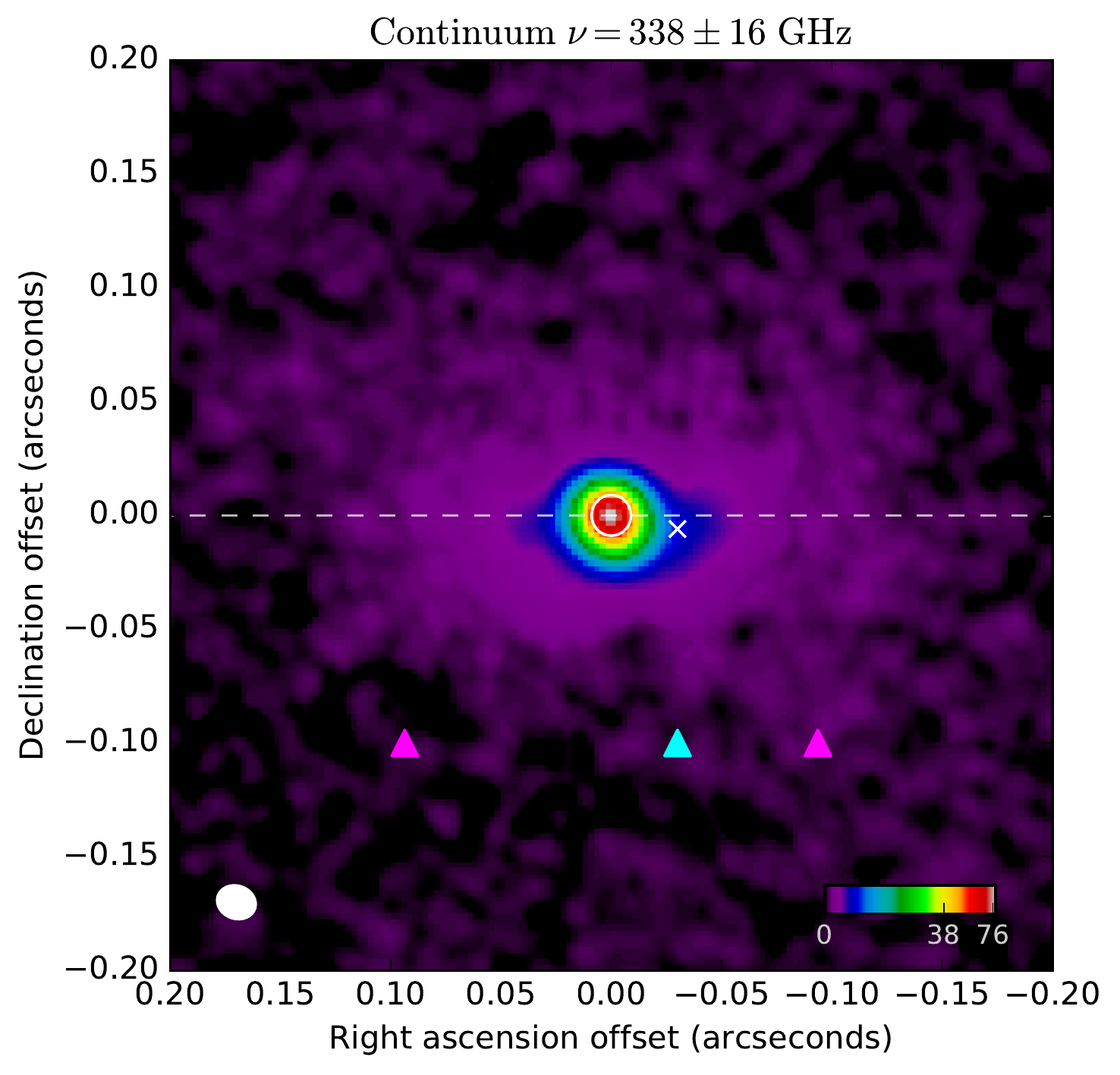}\hspace{5mm}
        \includegraphics[width=8cm]{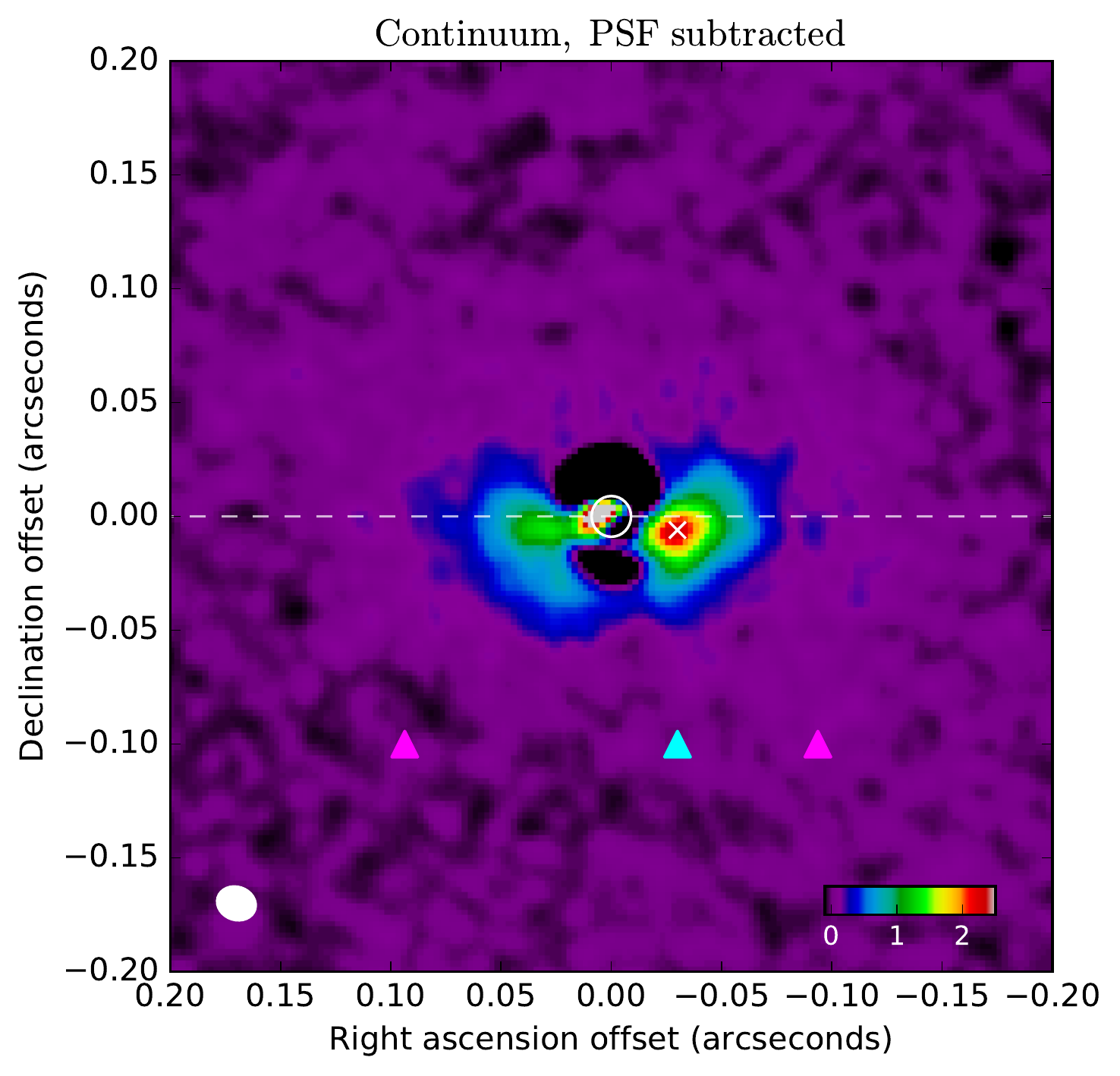}
        \caption{{\it Left panel:} Continuum emission from L$_2$\,Pup at $\nu = 338 \pm 16$\,GHz. {\it Right panel:} PSF-subtracted continuum emission map. The color scales are a function of the square root of the intensity in mJy\,beam$^{-1}$\,km\,s$^{-1}$.
        The white cross marks the position of the secondary source in the PSF-subtracted ALMA image.
        The triangular symbols are pointers to the inner rim of the dust disk (6\,AU, magenta) and of the secondary source detected in the visible by \citetads{2015A&A...578A..77K} (cyan), shifted by $-0.1\arcsec$ in declination.
        The position of source B from ALMA is marked with a white cross, and the size of the photosphere of the star is shown as a white circle.
        The $17.7 \times 14.5$\,mas beam is represented in the lower left corner of the images.
        \label{continuummap}}
\end{figure*}

The ALMA continuum emission map at $\nu = 338 \pm 16$\,GHz (Fig.~\ref{continuummap}, left panel) shows thermal emission from the AGB star and the dust disk at a resolution corresponding to 0.9\,AU at L$_2$\,Pup.
The east-west elongation of the diffuse emission is consistent with the major axis of the dust disk observed by \citetads{2014A&A...564A..88K}, \citetads{2015A&A...576A..46L,2015A&A...581C...2L}, \citetads{2015A&A...581A.127O} and \citetads{2015A&A...578A..77K} at infrared and visible wavelengths.
The visible image of L$_2$\,Pup from \citetads{2015A&A...578A..77K} is shown at the same scale as the ALMA maps for comparison.

\begin{figure}[]
        \centering
        \includegraphics[width=8cm]{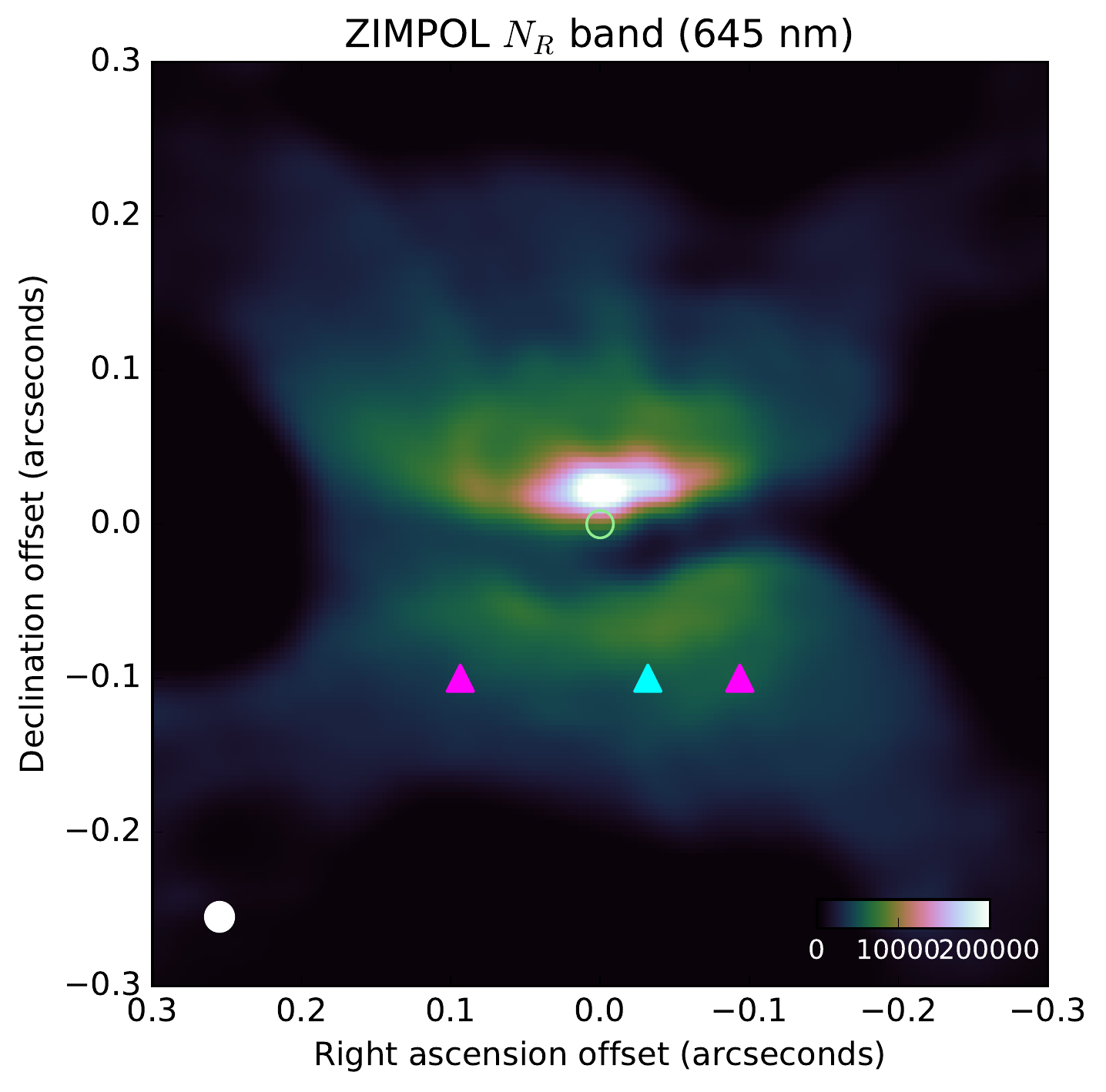}
        \caption{Visible image of L$_2$\,Pup from \citetads{2015A&A...578A..77K} for comparison with the ALMA maps, at the same scale.
        The beam size is represented by the ellipse in the lower left corner of the image.
        The radius of the inner rim (6\,AU) and the radius of the companion of L$_2$\,Pup are shown with magenta and cyan triangles, respectively.
        \label{ZIMPOL-image}}
\end{figure}

\subsubsection{Flux from L$_2$\,Pup~A}

The total flux density at the location of the AGB star including the contribution from the dust thermal emission is:
$f_A + f_\mathrm{dust} = 78.6 \pm 0.03\,\mathrm{mJy\,beam}^{-1}$.
We estimate the dust emission at the location of the AGB star from the mean continuum flux measured in the disk plane away from the stellar emission peak, as was done by \citetads{2014A&A...564A..88K} in the infrared.
We obtain $f_\mathrm{dust} = 1.4$\ mJy\,beam$^{-1}$, for which we adopt an arbitrary uncertainty of $\pm 0.3$\,mJy\,beam$^{-1}$ to account for possible variations in the local dust emission over the disk.
The flux from the L$_2$\,Pup~A is then
\begin{equation}
f_A = 77.2 \pm 0.3\ \mathrm{mJy\,beam}^{-1}.
\end{equation}
The angular diameter of the photosphere of L$_2$\,Pup~A has been measured by \citetads{2014A&A...564A..88K} using near-infrared interferometry at $\theta_\mathrm{LD} = 17.9 \pm 1.6$\,mas (see also \citeads{2015A&A...581A.127O}).
The major axis of the central source in the ALMA continuum map is 20.8\,mas at a position angle of $51^\circ$ ($N = 0^\circ$, $E = 90^\circ$, and the minor axis 18.9\,mas.
These values are slightly higher than the ALMA beam size ($17.7 \times 14.5$\,mas at $\mathrm{PA}=73^\circ$) and consistent with a partial resolution of the stellar photosphere. 

As a consistency check, we verified that the measured flux from the central AGB star $f_A$ is close to the emission of $f = 92$\,mJy expected for a blackbody at $T_\mathrm{eff} = 3500$\,K and with an angular diameter of $\theta = 18$\,mas.
\begin{figure}[]
        \centering
        \includegraphics[width=\hsize]{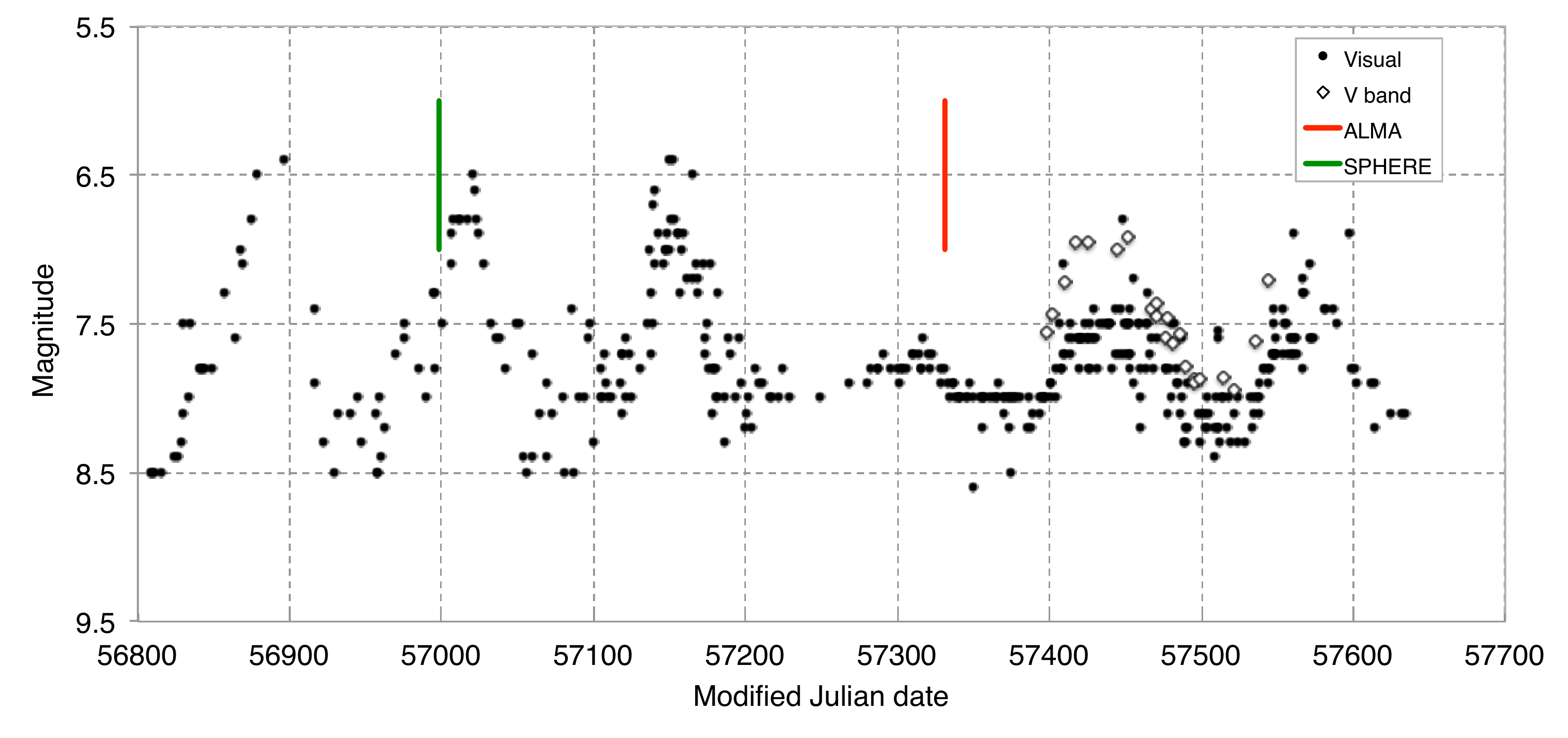}
        \caption{Photometric curve of L$_2$\,Pup in the visible from the AAVSO database, from 23 May 2014 to 2 September 2016. The vertical color segments represent the epochs of the SPHERE observations (green) by \citetads{2015A&A...578A..77K} and the present ALMA band 7 observation (red).
        \label{aavso}}
\end{figure}
The AAVSO light curve\footnote{\url{https://www.aavso.org}} of L$_2$\,Pup at visible wavelengths presented in Fig.~\ref{aavso} shows that our ALMA observations were obtained approximately one month after the expected maximum light, during the decreasing flux phase of the  cycle of L$_2$\,Pup.
During this phase, the star is cooler and fainter, which could explain part of the 20\% flux deficit compared to the expected value.
The photometric curve also shows that L$_2$\,Pup went through a particularly long minimum around $m_V = 8.0$ during the second half of 2015, including during our ALMA observations.
The deficit in flux could also be enhanced by this special minimum.
The small amplitude peak flux shortly before the ALMA epoch could indicate that the obscuration of the star by dust became particularly strong and compensated for the increase in flux of the AGB star.
The regular pulsational variability resumed in early 2016, although with a smaller amplitude than in 2014.

\subsubsection{Secondary source L$_2$\,Pup~B\label{psfsubtraction}}

As shown in Fig.~\ref{continuummap} (left panel), the west wing (to the right of the image) of the disk emission appears brighter than the east side.
To isolate this asymmetry, we estimated the contribution of the central star by fitting a two-dimensional elliptical Gaussian to the central source and we subtracted it from the continuum image.
The result of the PSF subtraction is presented in the right panel of Fig.~\ref{continuummap}.
The red wing of the disk (West) clearly generates significantly more flux than the blue wing.
A secondary source (hereafter source B, or L$_2$\,Pup~B) is observed in the subtracted image at a relative position of $\Delta \alpha = \deltalphaB \pm 2.5$\,mas and $\Delta \delta = \deltdeltaB \pm 2.5$\,mas with respect to the central object.
Source B appears superimposed on the emission from the disk, and is unresolved angularly.
The observed position corresponds to a projected separation of $\rho = \rhoabsB \pm \rhoBerr$\,mas, which is equivalent to a linear radius of $R_0 = \RB \pm \RBerr$\,AU. 
The relative position of source B with respect to A is closely coincident with that of the companion detected by \citetads{2015A&A...578A..77K} at visible wavelengths.

The flux density at the location of source B is $f = 2.33 \pm 0.05$\,mJy\,beam$^{-1}$.
The uncertainty takes into account the standard deviation of the background noise of the reconstructed image and of the PSF subtraction.
As for L$_2$\,Pup~A, the measured flux includes the emission from B plus the thermal emission of the surrounding dust.
We estimate the dust contribution from a point in the disk located symmetrically to B with respect to the central star, where we find $f_\mathrm{dust} = 1.34 \pm 0.10$\,mJy\,beam$^{-1}$.
The error bar includes a provision for a possible variability in the local dust emission in the disk.
The net flux from source B is then
\begin{equation}
f_B = 0.99 \pm 0.11\ \mathrm{mJy\,beam}^{-1},
\end{equation}
and the flux contrast between sources A and B at $\nu = 338$\,GHz ($\lambda = 887\,\mu$m) is
\begin{equation}
f_B / f_A = 1.28 \pm 0.13\%.
\end{equation}
This relative flux is considerable in absolute terms because of the very high brightness of the AGB star.
At visible wavelengths, the contribution of B is even higher in relative terms, as \citetads{2015A&A...578A..77K} found $f_B / f_A [V] = 24 \pm 10\%$ in the $V$ band ($\lambda = 554$\,nm) and $f_B/f_A [N_R] = 19 \pm 10\%$ in the $N_R$ filter ($\lambda = 646$\,nm).
The flux at visible wavelengths is most likely dominated by scattered light above the disk plane, that biases the color toward the blue (where scattering is more efficient).
Moreover, as a result of inhomogeneities in the dust disk, the flux of the central star and source B are potentially affected by differential absorption, which is particularly strong in the visible.


\subsection{Molecular emission \label{diskrotation}}

\begin{figure}[]
        \centering
        \includegraphics[width=8cm]{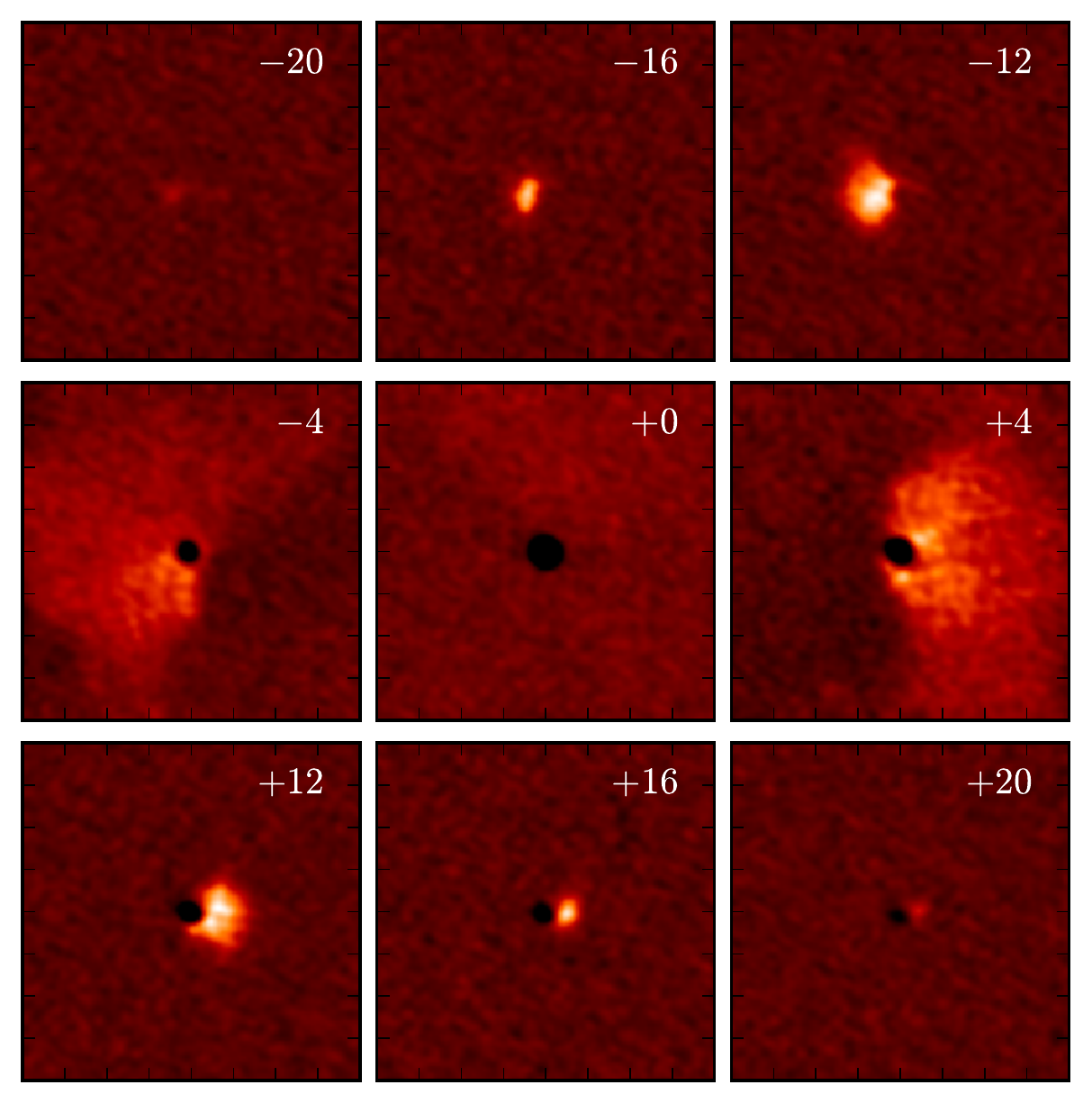}
        \caption{Continuum-subtracted channel maps of the $^{29}$SiO$(\varv=0,J=8-7)$ line.
        The field of view is $0.2\arcsec \times 0.2\arcsec$, and the indicated velocity is in km\ s$^{-1}$.
        \label{29SiO-channels}}
\end{figure}

We focus our analysis on the gas velocity field as derived from the $^{29}$SiO$(\varv=0,J=8-7)$ molecular line (Fig.~\ref{29SiO-channels}).
The molecular disk is centered on the star (which is visible in the continuum images presented in Fig.~\ref{continuummap}).
The central dark 'hole' is due to line absorption by the molecular gas located in front of the star.
We also present in Appendices~\ref{12COline} to \ref{SiSline} the images and position-velocity diagrams (hereafter PVDs) of the detected $^{12}$CO, $^{13}$CO, SO$_2$, SO and SiS lines that are also observed at high spectral resolution and high signal-to-noise ratio in the ALMA data.
A summary of the detected lines is presented in Table~\ref{l2pup-lines}.

\begin{table*}
        \caption{Molecular emission lines detected in L$_2$~Puppis.}
        \centering          
        \label{l2pup-lines}
        \begin{tabular}{lrcllr}
	\hline\hline
        \noalign{\smallskip}
        Molecule & \multicolumn{2}{c}{Quantum numbers}  & Sky freq. $\nu$ & Rest freq. $\nu_0$ & Upper-state \\
         & Vibrational & Rotational  & (GHz) & (GHz) & energy (K) \\
        \noalign{\smallskip}
        \hline    
        \noalign{\smallskip}
Unidentified & & & 330.437 & 330.477 \\
$^{13}$CO & $\varv=0$ & $3-2$ & 330.552 & 330.588 & 31.732 \\
H$_2$O & $\varv_2=2$ & $3(2,1) - 4(1,4)$ & 331.08 & 331.12 \\
$^{30}$SiO ? & $\varv = 3$ & $8-7$ & 331.93 & 331.97 \\
%
Unidentified & & & 331.99 & 332.03 \\
%
SO$_2$ & $\varv=0$ & $21(2,20)-21(1,21)$ & 332.055 & 332.091 & 219.525 \\
SO$_2$ & $\varv=0$ & $34(3,31)-34(2,32)$ & 342.724 & 342.762 & 581.919 \\
$^{29}$SiO & $\varv=0$ & $8-7$ & 342.943 & 342.981 & 74.077 \\
SiS & $\varv=1$ & $19-18$ & 343.063 & 343.101 & 1235.828 \\
SO $3\Sigma$ & $\varv=0$ & $8(8)-7(7)$ & 344.272 & 344.311 & 87.482 \\
$^{12}$CO & $\varv=0$ & $3-2$ & 345.758 & 345.796 & 33.192 \\
        \hline                      
        \end{tabular}
\end{table*}

\subsubsection{Position velocity diagrams}

\begin{figure}[]
	\centering
        \includegraphics[width=7cm]{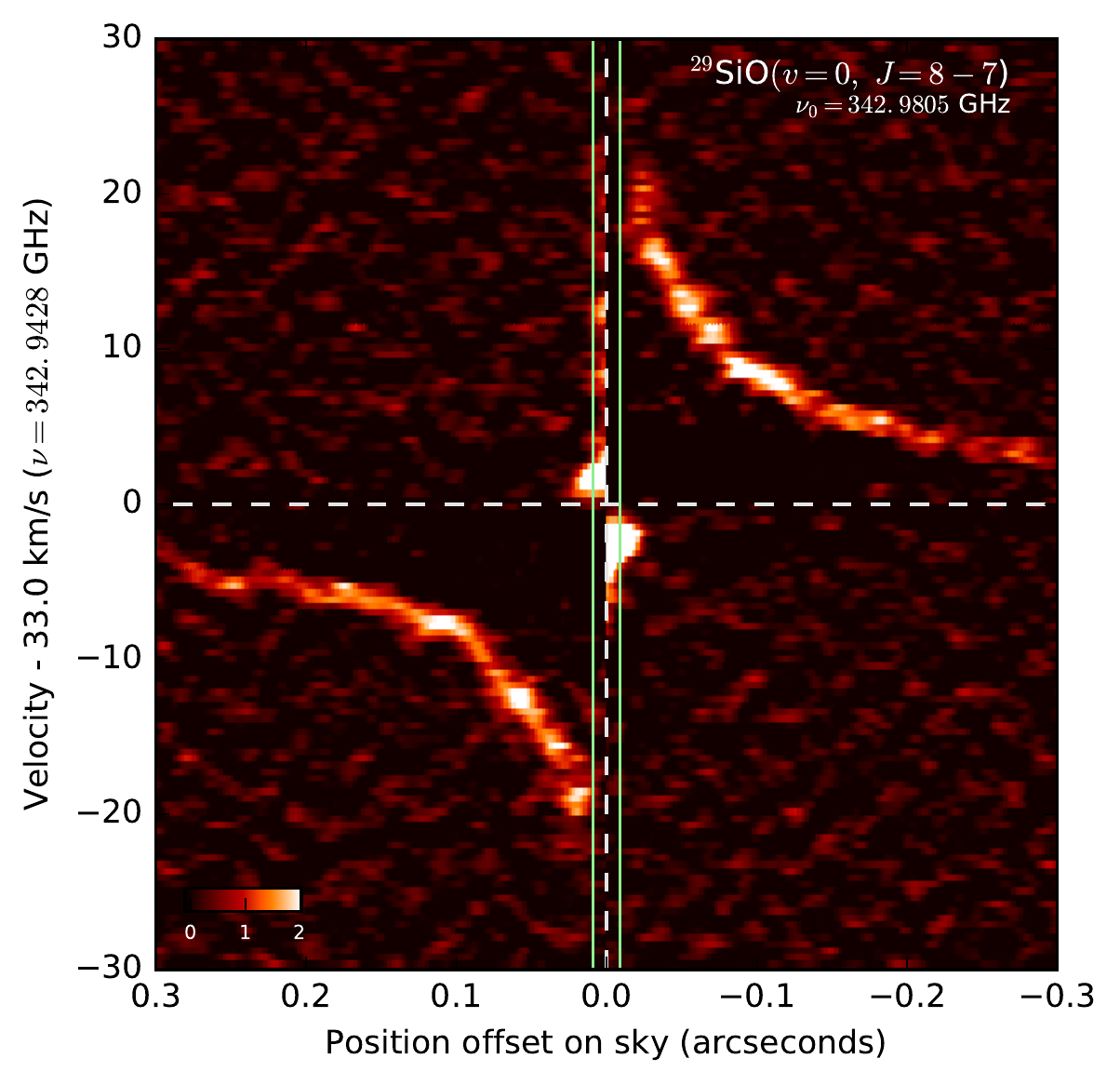}
        \caption{Derivative along position of the PVD of the $^{29}$SiO$(\varv=0,J=8-7)$ line.
	 The color scale in Jy\ arcsec$^{-1}$. The corresponding PVD is presented in Fig.~\ref{29SiO-pvd}.
        \label{29SiO-diffpv}}
\end{figure}

We computed PVDs for each emission line from the reconstructed ALMA image cubes using a virtual slit of 20\,mas oriented in the east-west direction (position angle $PA = 90^\circ$ with respect to $\mathrm{north}  = 0^\circ$), that is, along the plane of the circumstellar disk.
To constrain the value of the central mass from the orbital motion of the circumstellar gas, we estimated the maximum velocity as a function of the radius.
Several techniques have been employed in the literature to derive masses from PVD velocities (e.g.~\citeads{2016MNRAS.459.1892S,2013A&A...560A.103M}).
We adopted the innovative approach to compute the derivative of the PVD along the radial direction to determine the radius of maximum variation slope.
Compared to the use of a fixed flux detection threshold, for instance, this technique has the advantage to be less sensitive to the angular resolution of the observation (as long as the disk itself remains resolved).
An example of derivative of the PVD along the position axis is presented in Fig.~\ref{29SiO-diffpv} for the $^{29}$SiO$(\varv=0,J=8-7)$ line (see also Fig.~\ref{29SiO-pvd} for the corresponding PVD).
The edge of the velocity profile is clearly defined and easily measurable.

Using this technique, the determination of the spatial position of the point of inflection of the PVD is largely insensitive to biases.
Whether Keplerian or non-Keplerian, the pure rotation profile of the gas itself is theoretically sharp (i.e. a Heaviside function). 
There may be an intrinsic dispersion of the velocity in the disk due to turbulence, for example.
In this case, we would expect for the velocity profile to observe a convolution of the Heaviside function by a Gaussian-like function, and a broadening of the profile extension in radius for a given velocity.
But this operation would not change the position of the point of inflection, as the convolution of the Heaviside function already smoothed by a Gaussian by a second Gaussian (the angular resolution of the array) would not shift the point of inflection.
It would only reduce the maximum slope, hence the accuracy of the determination of its position.
This measurement technique is valid when the instrumental beam is smaller than the true width of the profile itself in a given spectral channel.
As a consequence, it does not work properly for the highest velocities, which exhibit a point-like flux distribution on the images and are located very close to the star.
The inner cavity (within 6\,AU from the star), however, is approximately $0.1\arcsec$ in radius, and therefore very well resolved by ALMA.


\subsubsection{Mass of the central object\label{massA}}

\begin{figure*}[]
        \centering
        \includegraphics[width=8cm]{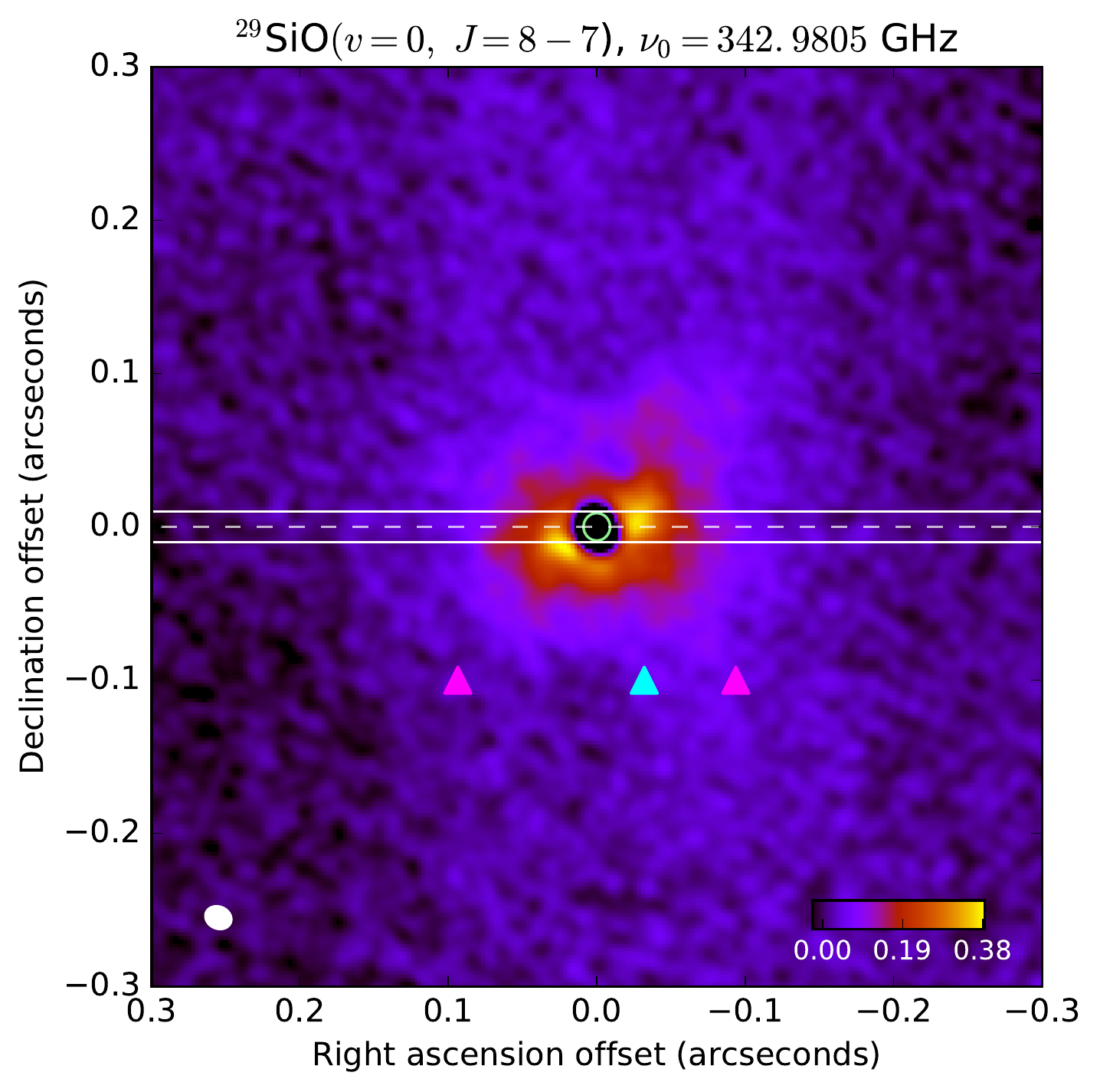} 
        \includegraphics[width=8cm]{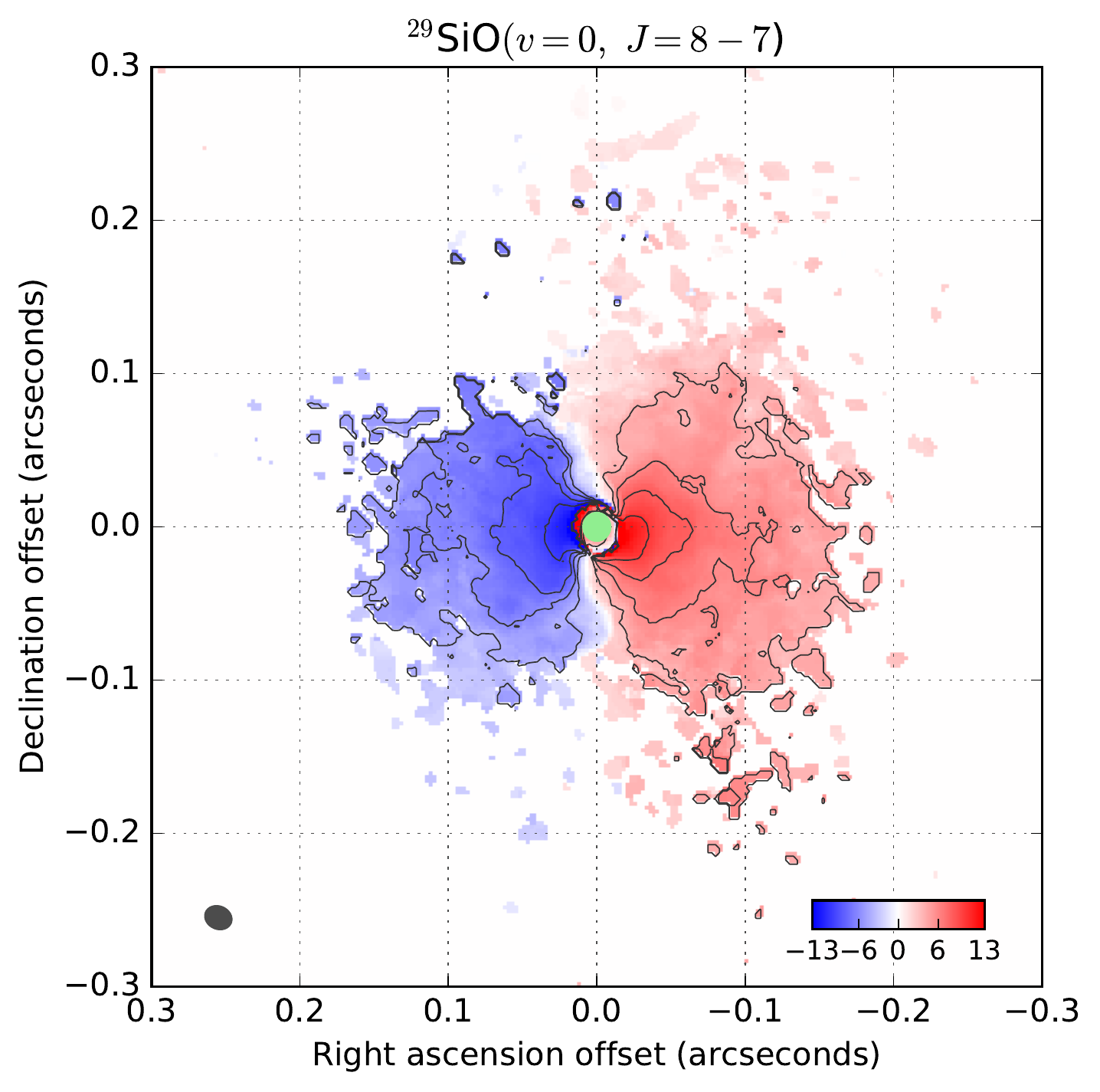}
        \caption{{\it Left panel:} Map of the emission from L$_2$\,Pup in the $^{29}$SiO$(\varv=0,J=8-7)$ line (left), integrated between radial velocities $-30$ and $+30$\,km\ s$ ^{-1}$ around the systemic velocity ($v_0 = 33$\,km\ s$ ^{-1}$).
        The pseudo-slit used to compute the position-velocity diagrams is represented with solid white lines.
        The linear color scale in Jy\ beam$^{-1}$\ km\ s$^{-1}$ is shown in the bottom right corner, and the beam size is represented by the ellipse in the lower left corner of the image.
        The radius of the inner rim (6\,AU) and the radius of the companion of L$_2$\,Pup are shown with magenta and cyan triangles, respectively.
        {\it Right panel:}  Map of the first moment emission of the velocity (the color scale is in km\ s$^{-1}$).
        The contours are drawn between 4 and 10\,km\,s$^{-1}$ with a 2\,km\,s$^{-1}$ step, and the size of the photosphere is shown with a light green disk.
        \label{29SiO-image}}
\end{figure*}

\begin{figure*}[]
        \centering
        \includegraphics[width=16cm]{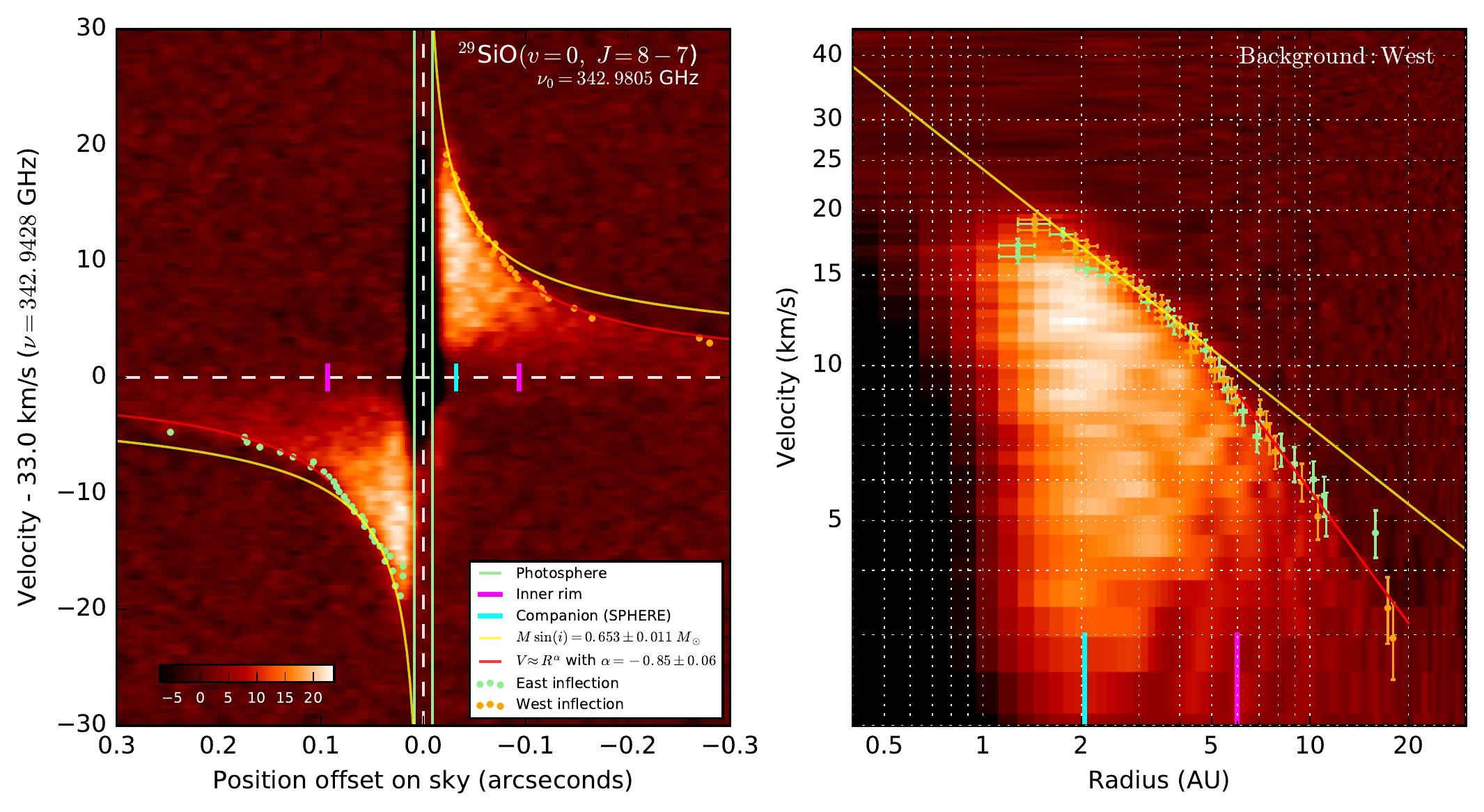}
        \caption{Position-velocity diagram with Cartesian coordinate axes (left) and with logarithmic coordinates (right) to show the Keplerian and non-Keplerian domains of the disk rotation velocity profile more clearly.
        The yellow curve is a Keplerian fit to the inner cavity, and the red curve is a power law fit to the non-Keplerian radius domain.
	 The color scale is in mJy\ beam$^{-1}$.
        \label{29SiO-pvd}}
\end{figure*}

The first moment of the $^{29}$SiO$(\varv=0,J=8-7)$ line velocity (Fig.~\ref{29SiO-image}) and the PVD (Fig.~\ref{29SiO-pvd}) point unambiguously at a rotating gaseous disk \citepads{2015A&A...579A.118H}.
The point of inflection of the PVD along the position axis for each velocity bin is represented with a light green or orange dot for the east and west sides of the disk, respectively.
They correspond to the peak positions visible in Fig.~\ref{29SiO-diffpv}.
The size of the stellar photosphere ($\theta_\mathrm{LD} = 17.9 \pm 1.6$\,mas; \citeads{2014A&A...564A..88K}) is shown with light green lines.
Color markings indicate the positions of the inner rim of the dust disk (magenta) and of the companion of L$_2$\,Pup (cyan) as observed by \citetads{2015A&A...578A..77K}.
The yellow curve in Fig.~\ref{29SiO-pvd} represents the best-fit Keplerian velocity profile to the inflection points, adjusted between radii of 2.5 and 4.5\,AU.
The velocity field over this radius domain is Keplerian as shown in the right panel of Fig.~\ref{29SiO-pvd}, with a power law dependence of the velocity in $R^{-1/2}$.
The radial position uncertainty in the point of inflection is fixed in the fit at $\pm 2.5$\,mas (one reconstructed image pixel), and the velocity uncertainty to $\pm 0.5$\,km\,s$^{-1}$.
This fit gives us the total mass enclosed within a radius of 2.5\,AU,
\begin{equation}\label{mproj}
m_\mathrm{2.5\,AU} \sin(i) = \masssini \pm \masssinierrstat \pm \masssinierrparallax\ M_\odot,
\end{equation}
where $i$ is the inclination of the rotating Keplerian disk on the line of sight.
The determined mass is linearly proportional to the parallax. The corresponding systematic uncertainty ($\pm 6.3\%$) dominates the error budget and is listed separately in Eq.~\ref{mproj}.
It should be noted that this mass estimate includes any additional mass contributor within the inner radius.
This implies that the mass contribution of the secondary source L$_2$\,Pup~B presented in Sect.~\ref{continuum} is included in this mass estimate.
However, we have essentially an upper limit for the mass of this secondary source (Sect.~\ref{massB}) that could be very low.
We therefore did not correct for its contribution and refer in the following to the mass $m_\mathrm{2.5\,AU}$ as the mass of L$_2$\,Pup~A ($m_A$).

The position angle of the gaseous disk rotation plane is oriented almost perfectly east-west, within $\pm 3^\circ$.
We determined this angle through a maximization of the PVD velocity amplitude, but the exact position angle within this range has a negligible influence on the determined mass.
The inclination $i$ of the circumstellar disk of L$_2$\,Pup on the line of sight has been estimated by \citetads{2014A&A...564A..88K} to $i = 84^\circ$ and slightly revised by \citetads{2015A&A...578A..77K} to $i = 82^\circ$.
\citetads{2015A&A...581A.127O} also concluded a high inclination from the partial obscuration of the stellar disk by the edge of the dust disk.
We therefore adopt a value of $i = 82 \pm 5^\circ$., and obtain the mass enclosed within 2.5\,AU in radius, identified as the mass of L$_2$\,Pup~A,
\begin{equation}
m_A = \mass \pm \masserrtot\ M_\odot\ (\pm \masserrpercent).
\end{equation}

\subsubsection{Sub-Keplerian disk rotation}

Figure~\ref{29SiO-pvd} shows that the rotation velocity of the gaseous disk changes from the purely Keplerian regime in the central cavity ($v \approx R^{-1/2}$) to a markedly sub-Keplerian regime beyond the inner rim of the dust disk (6\,AU).
The adjustment of a power law to the observed profile in the sub-Keplerian domain ($6 < R < 20$\,AU) gives the following radial dependence of the orbital velocity:
\begin{equation}
v = (\subkepA \pm \subkepAerr) \times R^{(\subkealpha \pm \subkepalphaerr)} \ \mathrm{km}\ \mathrm{s}^{-1}
\end{equation}
with $R$ expressed in astronomical units.
This velocity profile is represented with a red curve in Fig.~\ref{29SiO-pvd}.
We qualitatively interpret the deceleration beyond 6\,AU as being caused by the friction of the gas with the sub-Keplerian dust in the disk.
The dust grains are subject to the strong radiative pressure from the central star's radiation, and the effect of this radial force on the dust grains is equivalent to a reduction of the effective gravity.
This results in a reduced orbital velocity at a given radius compared to the Keplerian regime.
As the inner cavity has a low dust content, this effect is absent below a radius of $R \approx 5$\,AU, justifying our choice of the radial domain from 2.5 to 4.5\,AU (where the rotation is Keplerian) for the determination of the mass of the central object.
A detailed discussion of the sub-Keplerian rotation profile and structure of the disk is beyond the scope of the present work, and it will be presented in a forthcoming paper.

\subsubsection{Molecular emission from L$_2$\,Pup~B\label{molecB}}

\begin{figure*}[]
        \centering
        \includegraphics[width=5.5cm]{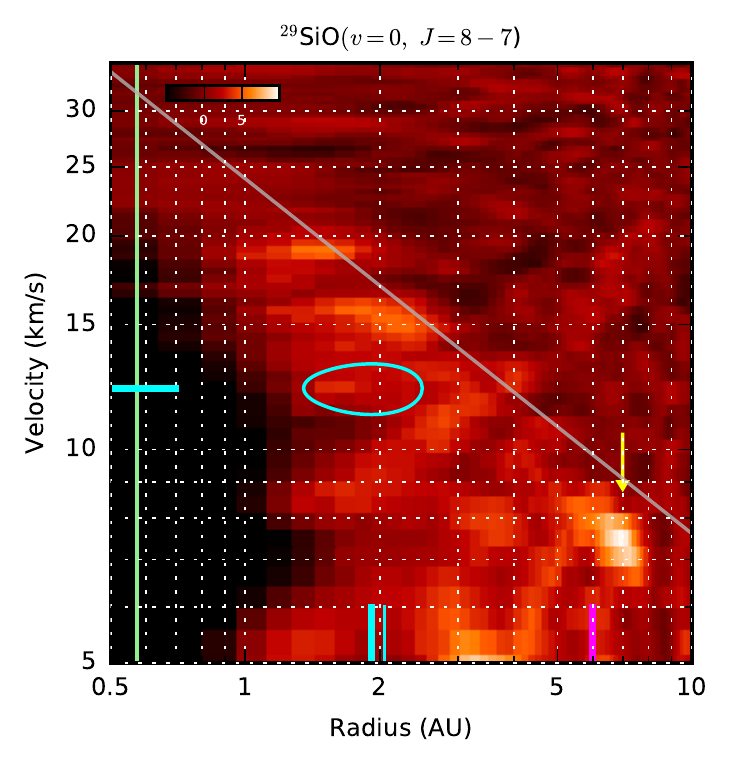} \hspace{2mm}
        \includegraphics[width=5.5cm]{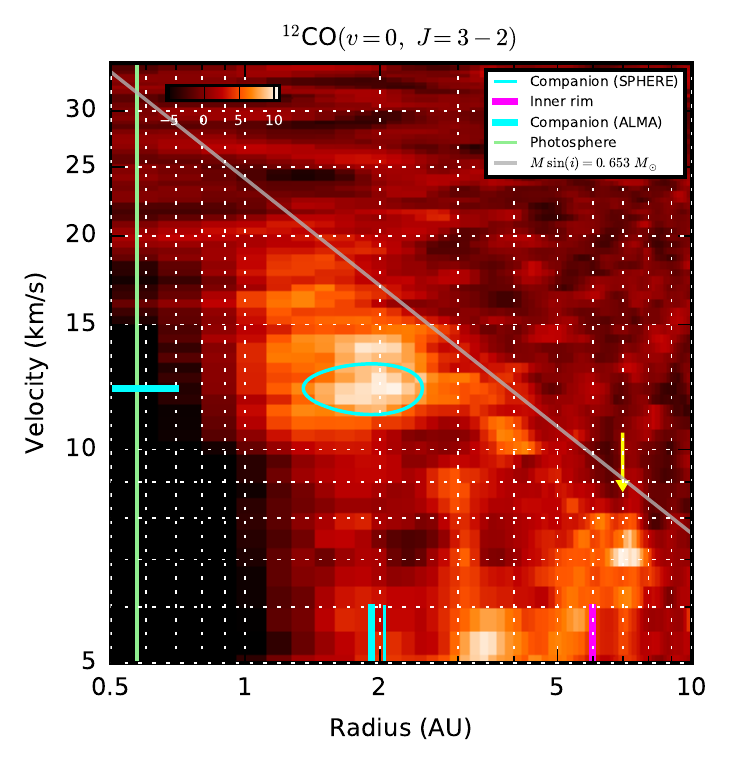} \hspace{2mm}
        \includegraphics[width=5.5cm]{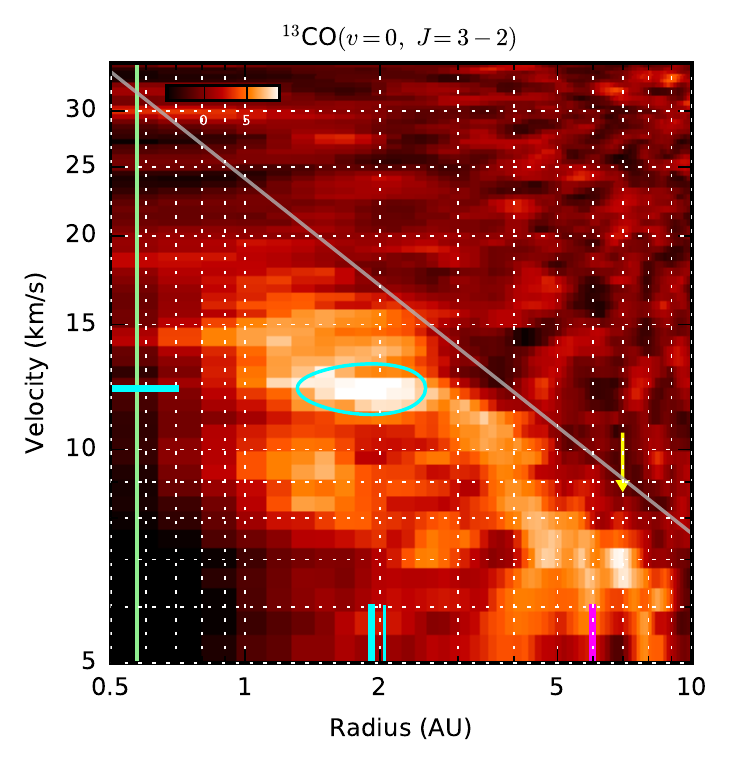}
        \caption{Position-velocity diagrams of the $^{29}$SiO$(\varv=0,J=8-7)$, $^{12}$CO$(\varv=0,J=3-2)$ and $^{13}$CO$(\varv=0,J=3-2)$ lines for the west part of the gaseous disk from which the east part of the PVD has been subtracted.
        The cyan ellipse is centered on the position of B, with a radial extension corresponding to the beam size and a velocity range of $12.2 \pm 1.0$\,km\,s$^{-1}$.
        The yellow arrow indicates the position of the notch at 7\,AU in the gaseous disk.
        The color scale is in mJy\,beam$^{-1}$.
        \label{COsubmaps}}
\end{figure*}

Figure~\ref{COsubmaps} shows the result of subtracting the eastern part of the PVD from the western part for the $^{29}$SiO, $^{12}$CO and $^{13}$CO lines.
We observe an excess molecular emission at the location of source B in the PVDs of $^{12}$CO and $^{13}$CO, but not in that of $^{29}SiO$.
We applied a global multiplicative factor of 0.8 to the east PVD as the emission is stronger in average (possibly due to absorption from dust), and we smoothed it using a moving median box of $7.5\,\mathrm{mas} \times 1.2\,\mathrm{km\,s}^{-1}$ to improve the signal-to-noise ratio.
The emission at the radius of source B is observed at a radial velocity of $v_0 = +12.2 \pm 1.0$\,km\,s$^{-1}$, represented with a blue ellipse in Fig.~\ref{COsubmaps}.
As a remark, an excess emission is also present in the $^{29}$SiO PVD (Fig.~\ref{29SiO-pvd}) at the same radial velocity, but it is essentially symmetrical between the east and west wings and therefore cancels out in the subtraction.
We also observe an excess emission at the radius of source B and around 12\,km\ s$^{-1}$ velocity in the PVD of the SO$\ 3\Sigma\ (\varv=0,8(8)-7(7))$ that is presented in Appendix~\ref{SOline}.
The frequency coverage of the opposite wing of the disk is incomplete, however, and does not permit the same subtraction as for the other lines.

A compact emission is also identified at a radius of 7\,AU and a velocity of $7.0 \pm 0.5$\,km\ s$^{-1}$ in the three subtracted PVDs (yellow arrows in Fig.~\ref{COsubmaps}).
In contrast to the emission identified at 2\,AU, it corresponds to a deficit of the east part of the PVD (i.e.~opposite to the companion) rather than an excess emission in the west.
This notch is visible in the non-subtracted PVDs (e.g.~Fig.~\ref{29SiO-pvd}) in the east, at a radius of 7\,AU, that is, slightly beyond the inner rim of the dust disk.
The origin of this notch is unknown, but it could indicate a furrow in the disk.
It is located at a similar radius as the origin of plume \#1 observed by \citetads{2015A&A...578A..77K} (see also Sect.~\ref{nebfeatures}).

\section{Discussion \label{discussion}}

\subsection{Mass and evolutionary state of L$_2$\,Pup~A\label{massdiscussion}}

\begin{figure}[h!]
        \centering
        \includegraphics[width=\hsize]{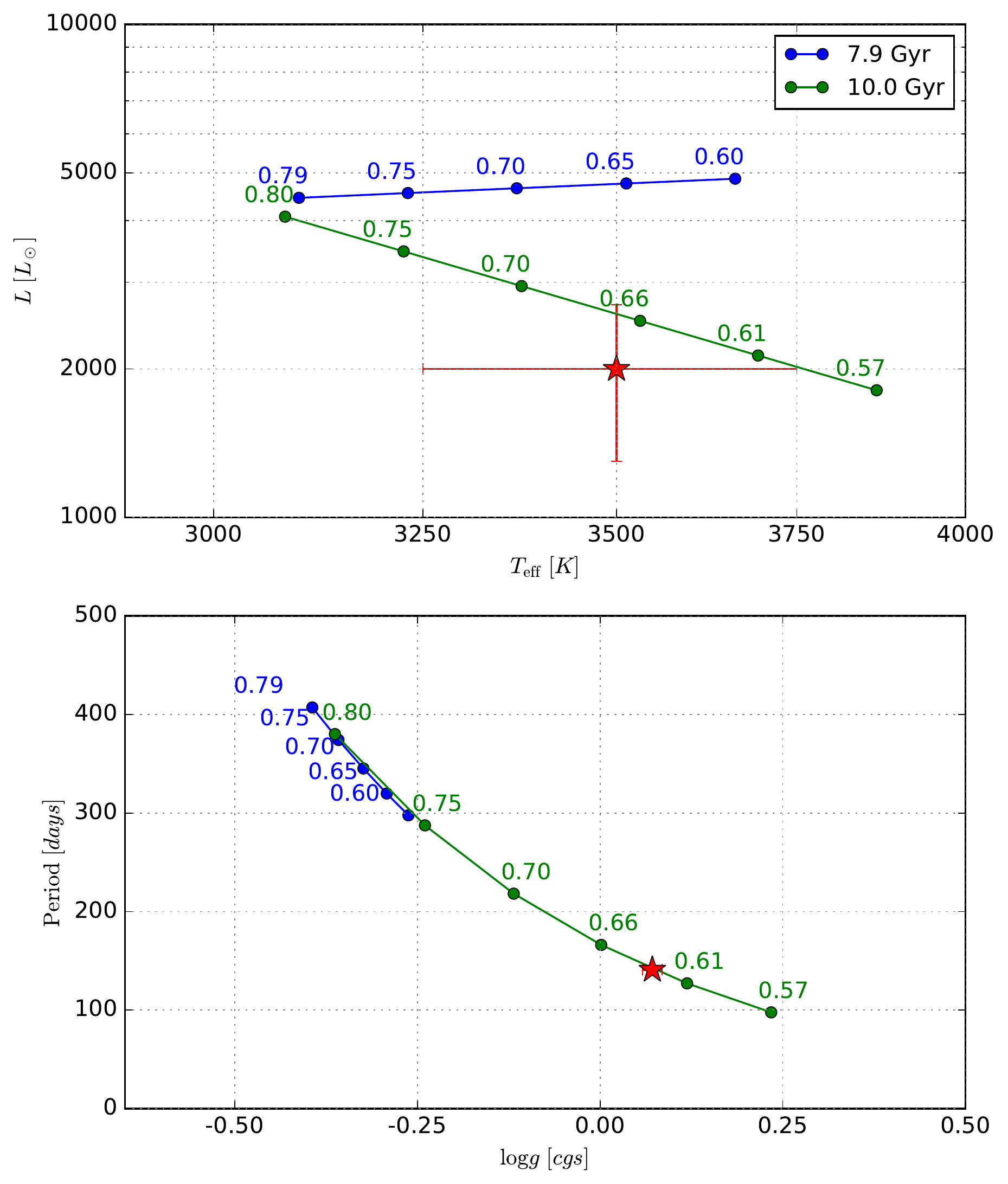}
        \caption{Position of L$_2$\,Pup compared to isochrones from the evolutionary models by \citetads{2008A&A...482..883M, 2013MNRAS.434..488M}.
        The mass of the star at the age of the isochrone is indicated as labels of each curve.
        {\it Top panel:}  Hertzsprung-Russell diagram.
        {\it Bottom panel:} Pulsation period vs.~$\log g$.
        \label{evomodels}}
\end{figure}

We compared the measured parameters of the star listed in Table~\ref{l2pupparams} (left columns) to the database of PARSEC+COLIBRI models of thermally pulsating AGB stars (TP-AGB) developed by \citetads{2008A&A...482..883M, 2013MNRAS.434..488M} in version PR16 \citepads{2016ApJ...822...73R}.
The pre-AGB evolution was computed using the PARSEC code \citepads{2015MNRAS.452.1068C}.
Based on its Galactic space velocity, \citetads{2002ApJ...569..964J} proposed that L$_2$\,Pup belongs to the thick-disk population of the Galaxy (i.e.~is a low-metallicity star).
This is consistent with the low-mass we measure, and we therefore adopted a sub-solar metallicity of $Z=0.008$.
We retrieved the isochrones from the CMD 2.8 web site\footnote{\url{http://stev.oapd.inaf.it/cgi-bin/cmd}}.
We chose to not interpolate them as the correlations between parameters are potentially complex.
A comparison of selected L$_2$\,Pup parameters compared to the COLIBRI isochrone predictions is presented in Fig.~\ref{evomodels}.

The closest COLIBRI model to the observed properties of the star has the parameters listed in Table~\ref{l2pupparams} (right columns).
Thanks to our knowledge of the mass and pulsation period, the model properties are well constrained and non-degenerate.
The agreement between the model and observed parameters is very good.
The measured effective gravity ($\log g$) was derived from the angular diameter measured by  \citetads{2014A&A...564A..88K} using near-infrared interferometry, combined with the \emph{Hipparcos} parallax and the mass $m_A$ determined in the present work.
The difference in $\log g$ with the observed value is well within the error bar, although no correction was made for the pulsation phase of the star.
The pulsation period predicted by the model (141.5\,days) is very close to the observed period ($138 \pm 1.7$\,days).
As expected for this evolved star, the fraction of helium in the core is very high, close to 90\% in mass.
%
The mass loss predicted by the COLIBRI model $\dot{M} = 3 \times 10^{-8}\,M_\odot\ \mathrm{yr}^{-1}$ is lower than the values estimated by \citetads{2002A&A...388..609W}, \citetads{2002MNRAS.337...79B} and \citetads{2002ApJ...569..964J} by about an order of magnitude.
This parameter is particularly difficult to model, and the agreement is still relatively satisfactory.
To explain the discrepancy, we could speculate that the mass loss of L$_2$\,Pup is currently enhanced compared to its medium-term average value (over a millenium, for example).
In this context, \citetads{2016MNRAS.460.4182C} proposed that the formation of the dust disk could be the result of a recent, maybe still ongoing, enhanced mass-loss event.
The influence of the companion on the mass loss is also uncertain.
The lifetime of L$_2$\,Pup in the TP-AGB phase is expected to be only on the order of 0.5\,Myr \citepads{2014ApJ...790...22R} which is very short and makes the proximity of this star particularly remarkable from a statistical point of view.

\begin{table*}
        \caption{Observed properties of L$_2$\,Pup and best COLIBRI model parameters.\label{l2pupparams}}
        \centering          
        \begin{tabular}{llclc}
	\hline\hline
        \noalign{\smallskip}
        Parameter & Observed  & Ref. & Model & $\Delta[\sigma]$$^a$ \\
        \noalign{\smallskip}
        \hline
        \noalign{\smallskip}
	Current mass $M_\mathrm{act}\ [M_\odot]$ &  $\mass \pm \masserrtot\ $ & K16 & $0.63$ & $-0.6$ \\
	Initial mass $M_\mathrm{ini}$ &  &  & $0.98\ M_\odot$ & - \\
	Radius $R\ [R_\odot]$ & $\radius \pm \radiuserr$ & K14 & $121$ & $-0.1$ \\
	Effective gravity $\log g$\,[cgs] & $\logg \pm \loggerr$ & K16 & $0.072$ & $+0.2$\\
	Luminosity $L\ [L_\odot]$ & $\luminosity \pm \luminosityerr$ & K14 & $2347$ & $+0.6$ \\
	Effective temperature $T_\mathrm{eff}$\ [K] & $\efftemp \pm \efftemperr$ & K14 & $3629$ & $+0.5$ \\
	Pulsation period $P$ [days] & $\period \pm \perioderr$ & B02 & $141.5$ & $+1.9$ \\
	Pulsation mode order &  &  & 0 & - \\
	Metallicity $Z$ &  &  & $0.008$ & - \\
	Carbon-to-oxygen ratio C/O &  &  & $0.46$ & - \\
	Mass of helium core $M_\mathrm{He}\ [M_\odot]$ &  &  & $0.55$ & - \\
	Mass-loss rate $\dot{M}\  [M_\odot\,\mathrm{yr}^{-1}]$ & $5\times10^{-7}$ & B02 & $3\times10^{-8}$ & - \\
        \hline
        \end{tabular}
	\tablefoot{$^a$$\Delta$ is the difference between the model and observed values expressed in number of times the error bar of the measurement.}
        \tablebib{
	B02: \citetads{2002MNRAS.337...79B}; K14: \citetads{2014A&A...564A..88K}; K16: present work
	}
\end{table*}

In summary, L$_2$\,Pup's current evolutionary state is a thermally pulsating AGB star, which is a brief stage of its evolution.
Its initial mass on the main sequence was very close to solar, and its age is approximately twice that of our star.
L$_2$\,Pup thus provides a remarkable analog of the Sun when it will enter the final phases of its evolution, shortly before metamorphosing from a red giant into a planetary nebula and becoming a white dwarf.

\subsection{Physical properties of L$_2$\,Pup~B \label{companion}}

\subsubsection{Mass from molecular disk dynamics \label{massB}}

The velocity profile of the disk and the position of the central AGB star allow us to estimate the mass of L$_2$\,Pup~B.
As the molecular disk revolves around the barycenter $G$ of the enclosed mass, we can constrain the mass of source B from the difference in position between the AGB star's photocenter $Ph$ (taken as a proxy of the center of mass of the star) and the geometrical  center of rotation $G$ of the gaseous disk.

We measured the position of $G$ by matching the Keplerian velocities on each side of the PVD (east and west) of the $^{29}$SiO$(\varv=0,J=8-7)$ line (Fig.~\ref{29SiO-pvd}).
Thanks to the very high signal-to-noise ratio, we measured the position of $G$ with an accuracy of $\pm 0.2$\,pix ($\pm 0.5$\,mas).
We shifted all images (line and continuum) to define this position as the zero of the relative coordinate grid.
The position $Ph$ of the photocenter of the AGB star is derived from a bidimensional Gaussian fit over the central part of the continuum emission.
The possibility exists that a position shift is introduced by an asymmetric masking of the stellar photosphere by the northern edge of the dust disk \citepads{2015A&A...581A.127O}.
This photocenter displacement is expected to be of smaller amplitude at millimeter wavelengths than in the visible or near-infrared because of the higher transparency of the dust, but we conservatively adopt a systematic uncertainty of $\pm 0.2$\,pix ($\pm 0.5$\,mas) on the measurement.

We therefore obtain a relative position in right ascension of $Ph$ with respect to $G$ of
\begin{equation}
\Delta \alpha[Ph-G] = +0.22 \pm 0.30\,\mathrm{pix} = +0.55 \pm 0.75\,\mathrm{mas}
\end{equation}
This difference is positive as $Ph$ is located slightly to the east of $G$.
The agreement between the positions of $G$ and $Ph$ is therefore very good and statistically compatible with zero.

Considering the observed position offset in right ascension $\Delta \alpha[BA] = +30.0 \pm 2.5$\,mas between L$_2$\,Pup~B and the AGB star, we conclude that the mass of B is
\begin{equation}
m_B = \frac{\Delta \alpha[Ph-G]}{\Delta \alpha[BA] - \Delta \alpha[Ph-G]} \times m_A
\end{equation}
We therefore derive
\begin{eqnarray}
m_B & = & 0.019 \pm 0.025 \times m_A \\
 & = & 0.012 \pm 0.016\,M_\odot \\
 & = & \massBjup \pm \massBjuperr\,M_\mathrm{Jup}.
 \end{eqnarray}
We assumed in this reasoning that only one companion source is present in addition to the central star A. The presence of other sources could induce a different combined shift of the barycenter that might bias the mass estimate.

\subsubsection{Orbital period and radius}

For simple geometrical reasons (see e.g.~Fig.~\ref{l2pupsketch}), the observed velocity $v_0$ of source B is sub-Keplerian at the observed projected separation $R_0$ and corresponds to the projected component of the orbital velocity $v$ of source B.
Its orbital radius $R_B$ is necessarily larger than its currently observed radius $R_0$ from A, and is given by the expression:
\begin{equation}
R_B = \left(\frac{R_0}{v_0}\,\sqrt{G\,m_A}\right)^{2/3}
\end{equation}
where $m_A = \mass \pm \masserrtot\ M_\odot$, $R_0 = 1.92 \pm 0.12\ R_\odot$ and $v_0 = 12.2 \pm 1.0$\,km\,s$^{-1}$.
We assumed that the orbit of B is circular and coplanar with the disk.
This may not be perfectly true in reality as the position of source B in the continuum appears slightly south of the plane of the disk.
The actual orbital plane of B may therefore be moderately tilted with respect to the plane of the disk.
Under the assumption that the molecular emission does come from source B, the orbital radius is therefore $R_B = 2.43 \pm 0.16 $\,AU corresponding to a maximum angular separation of $\rho_B = 38 \pm 3$\,mas.
The orbital velocity (circular orbit) is $v_\mathrm{orb} = 15.4 \pm 1.0$\,km\ s$^{-1}$.
From Kepler's third law, the orbital period is $P_\mathrm{orb} = 4.69 \pm 0.45$\,years.
As shown in Fig.~\ref{l2pupsketch}, this period is consistent with the small observed astrometric displacement of source B between the ZIMPOL (2014.93) and ALMA (2015.84) epochs that are separated by 0.91\,years.
The two epochs would then correspond to almost symmetric phases with respect to the maximum elongation ($-33^\circ$ and $+38^\circ$ for ZIMPOL and ALMA, respectively).
The ephemeris of the separation $\Delta \alpha_{[B-A]}$ in right ascension between source B and L$_2$\,Pup~A is therefore
\begin{equation}
\Delta \alpha_{[B-A]}(T) = -\rho_B \ \cos \left[ 2 \pi\ \phi(T) \right]
\end{equation}
with $\phi(T) =  (T - T_0)/P_\mathrm{orb}$, and the ephemeris of the radial component $v(T)$ of the orbital velocity is given by:
\begin{equation}
v(T) = v_\mathrm{orb} \ \cos \left[ 2 \pi\ \phi(T) \right]
\end{equation}
where $T$ is the observing date expressed as decimal year and $T_0 = 2015.1$ the epoch of maximum elongation to the west.
If the orbital motion ephemeris we determine is correct, the maximum elongation of source B to the east of L$_2$\,Pup~A should occur around June 2017.
The radial velocity would then be approximately $v_\mathrm{rad} = -15$\,km\,s$^{-1}$.

\subsection{Nature of L$_2$\,Pup~B}

Considering our uncertainty domain on the mass of L$_2$\,Pup~B ($m_B = \massBjup \pm \massBjuperr\,M_\mathrm{Jup}$), we can exclude that it is a very low-mass star at a $4\sigma$ level.
We are left with three hypotheses, ordered by increasing mass, that we critically discuss in the following paragraphs:
\begin{enumerate}
\item a dense, coreless aggregate of gas and dust,
\item a planet ($m_B \leq 12\,M_\mathrm{Jup}$),
\item a low-mass brown dwarf ($12\,M_\mathrm{Jup} < m_B < 30\,M_\mathrm{Jup}$).
\end{enumerate}
For simplification, we address the planet and low-mass brown dwarf hypotheses together, using the same ``compact body'' term.

\subsubsection{Continuum and molecular emission}

The continuum flux contribution of B is considerable in absolute terms ($f_B = 1.03 \pm 0.05\ \mathrm{mJy}$ assuming it is unresolved), and its molecular emission is also very strong in the CO lines.
These two characteristics can be explained by the emission of a dense aggregate of dust and gas, but the thermal emission of a planet or brown dwarf is clearly insufficient.
However, L$_2$\,Pup~B evolves in a particular environment, that is hot,
rich in molecules and possibly also contains refractory dust that condensed in the AGB star wind.
Assuming that B is a compact body, it is therefore located in a very favorable position to accrete material.
The emission from the accreted material, either from an extended atmosphere or an accretion disk, could strongly outshine the thermal emission from the planet itself.
An approximate computation using the formulae by \citetads{1971ARA&A...9..183P} shows that the Roche lobe of a $\massBjup\,M_\mathrm{Jup}$ companion located at 2.5\,AU from a $0.65\,M_\odot$ star has a radius of $r_\mathrm{Roche} = 0.3$\,AU.
%
A detailed model of the system is beyond the scope of the present work, but we speculate that the accretion of the AGB wind by source B fills the Roche lobe of the companion with a sufficient quantity of material to produce the observed continuum and line emission.
If confirmed, this configuration would provide valuable constraints for hydrodynamical models of star-planet interactions at the late stages of stellar evolution (see e.g.~\citeads{2016MNRAS.458..832S,2014ASPC..487..401W}).

We conclude that the high observed continuum and molecular emission of L$_2$\,Pup~B is compatible with the object being either a core-less aggregate or a compact body.

\subsubsection{Persistence over time}

L$_2$\,Pup B has been detected at both the ZIMPOL (2014.93, \citeads{2015A&A...578A..77K}) and ALMA (2015.84) epochs separated by 0.91\,years.
From hydrodynamical simulations, gas clumps in an idealized radial outflow typically do not survive for long periods of time, mainly because of the internal pressure gradient.
The predicted lifetimes are on the order of weeks or months (depending on the size of the clump, temperature gradient, density gradient, etc.).
The dust (both small and large grains) is almost unaffected, as it is not subject to internal pressure.
The only influence of the outflow on the dust is a gradual elongation of the clump perpendicular to the local radial velocity vector field.
In a rotating medium as observed around L$_2$\,Pup, the gas would not survive as a clump because of the intense shear originating from the Keplerian rotation, and would be quickly dissolved.
These predictions are difficult to reconcile with the observed  molecular emission enhancement in the CO lines (Sect.~\ref{molecB}).
Moreover, as a result of the strong radiative pressure, a fluffy clump of dusty material would most likely be quickly blown away from the star and incorporated in the dust disk.
Overall, the observed asymmetry between the west and east wings of the disk is difficult to explain by simple inhomogeneity of the disk structure, since the turbulence in, and the viscosity of the medium would conspire against its development.

The persistence of source B over a period of one year favors the hypothesis of a compact object over a core-less aggregate.

\subsubsection{Relation with the nebular features\label{nebfeatures}}

\begin{figure}[]
        \centering
        \includegraphics[width=4.4cm,page=1]{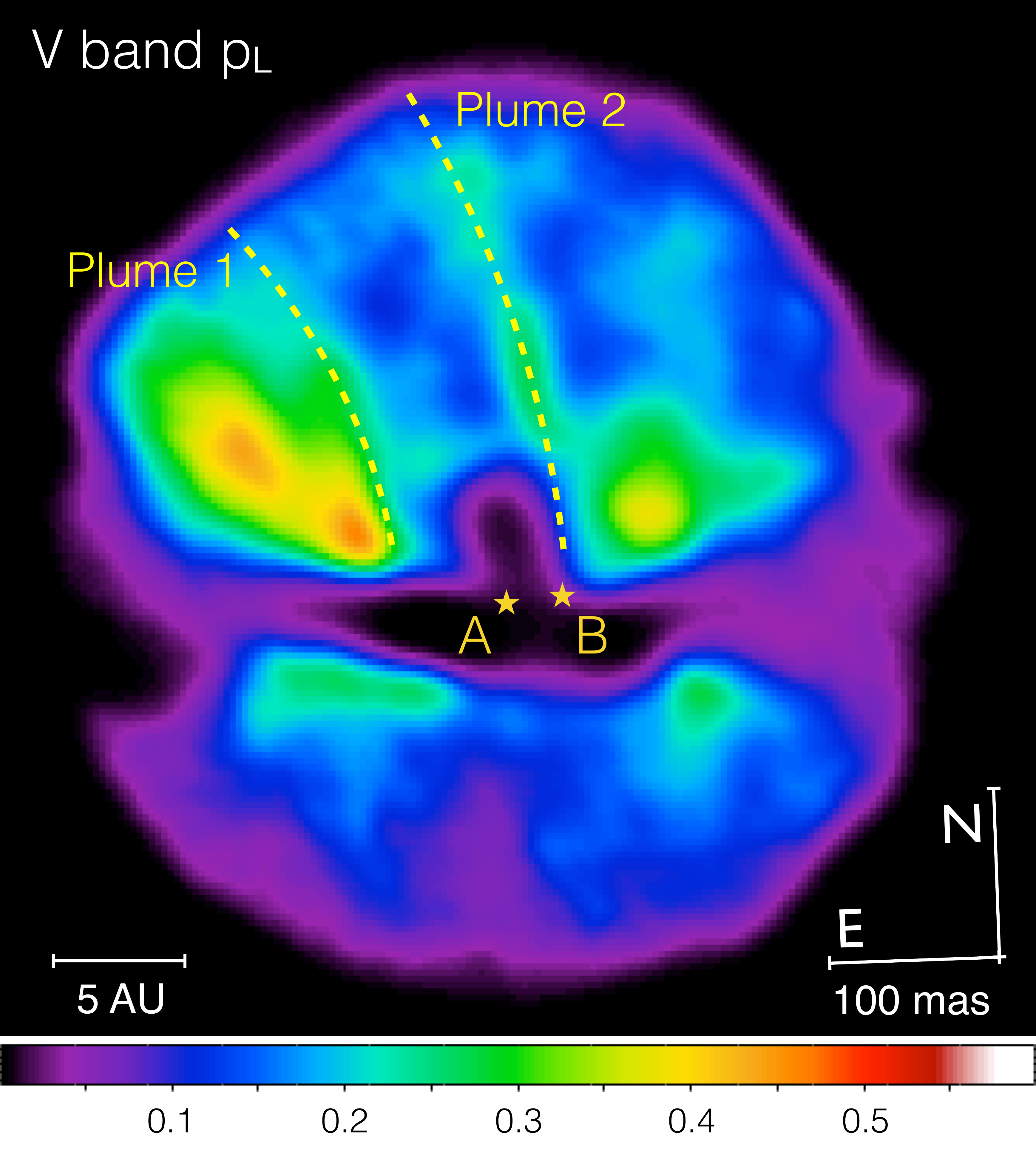}\hspace{1mm}
        \includegraphics[width=4.4cm,page=2]{Figures/Figures-L2Pup-SPHERE.pdf}
        \caption{{\it Left panel:} Map of the degree of polarization in the nebula of L$_2$\,Pup showing plume \#2 (from \citeads{2015A&A...578A..77K}).
        {\it Right panel:} Intensity image at $\lambda = 4.05\,\mu$m showing the loop (from \citeads{2014A&A...564A..88K}).
        The angular scale and orientation of the two panels are identical.
        \label{l2pupfeatures}}
\end{figure}

\begin{figure}[]
        \centering
        \includegraphics[width=\hsize,page=2]{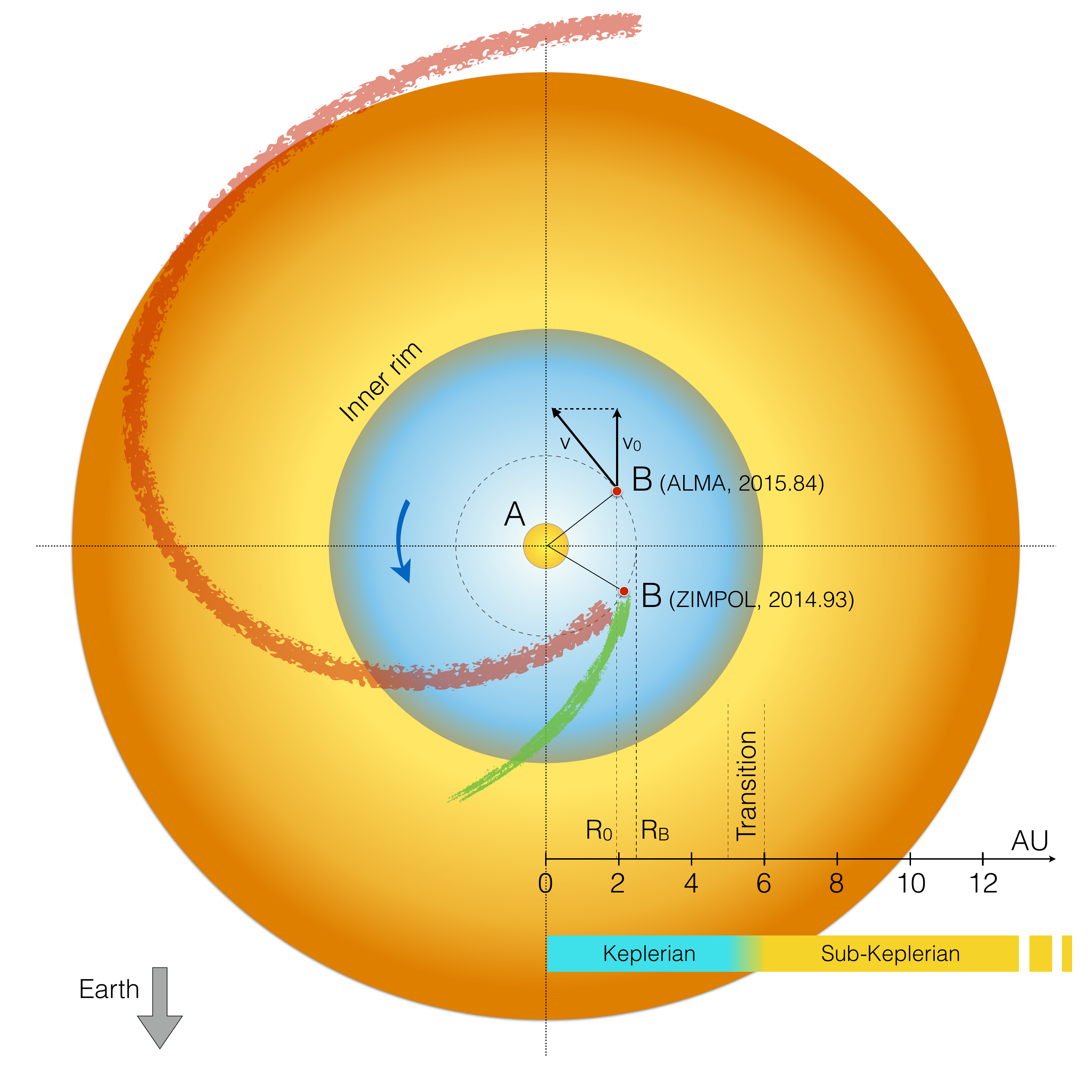}
        \includegraphics[width=\hsize,page=3]{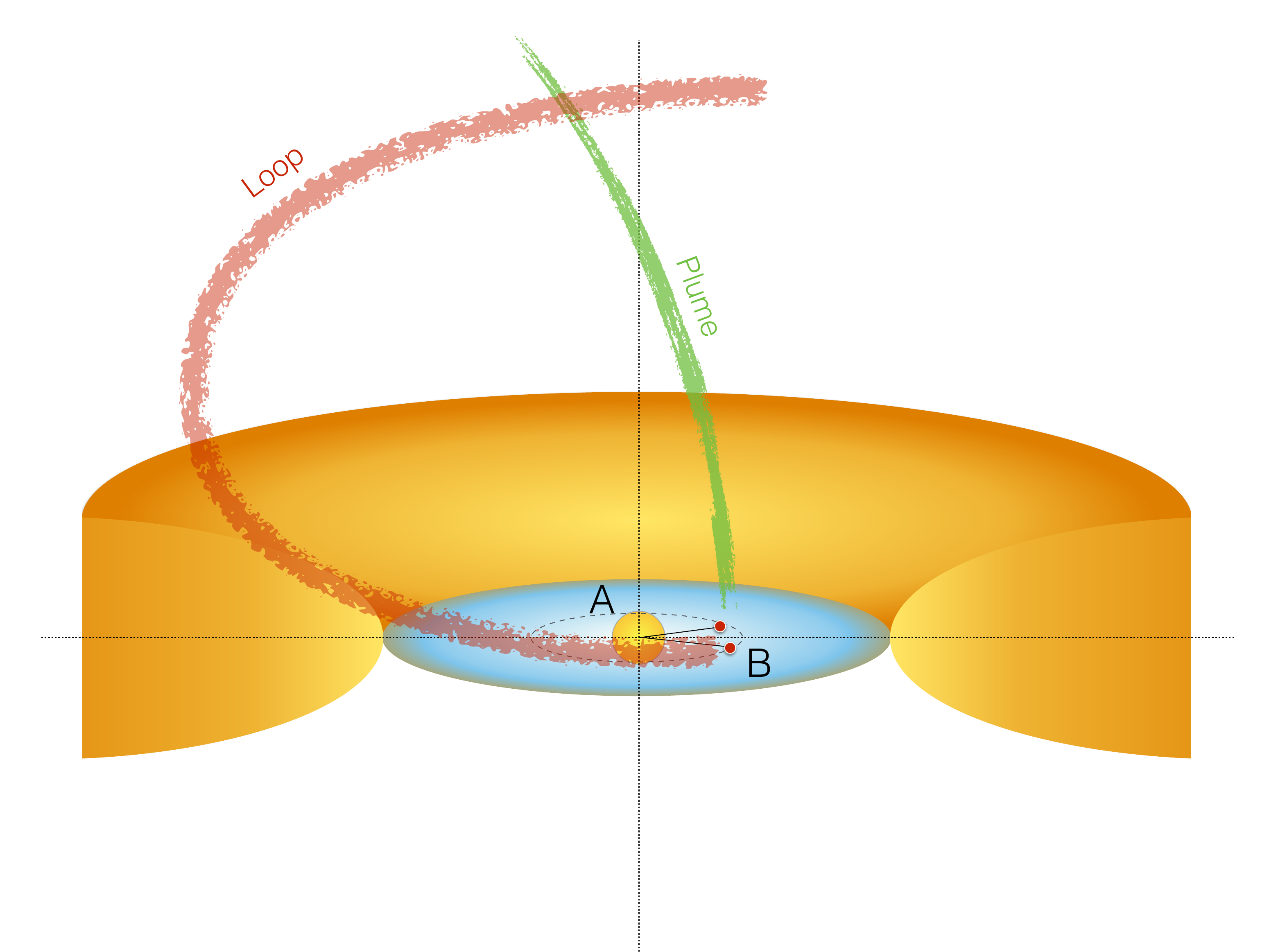}
        \caption{Schematic view of the environment of L$_2$\,Pup as seen from the north pole of the disk (upper panel) and close to the its plane (bottom panel). The dust component is shown in yellow-orange, and the median plane of the inner gaseous disk is represented in blue.
        The position of source B is represented for the ZIMPOL and ALMA epochs.
        The components are shown to scale.
        \label{l2pupsketch}}
\end{figure}

From visible polarimetric imaging, two plumes (labeled \#1 and \#2) were identified by \citetads{2015A&A...578A..77K} in the envelope of L$_2$\,Pup at visible wavelengths.
An extended loop was also detected in the thermal infrared ($\lambda = 4\,\mu$m), starting from the western wing of the disk and developing toward the northeast.
The location of these features is presented in Fig.~\ref{l2pupfeatures} (see also Fig.~\ref{ZIMPOL-image}).
Both plumes appear to originate in the plane of the disk, and the position of L$_2$\,Pup~B coincides with the origin of plume \#2.
Plume \#1 is less extended and more irregular than plume \#2, and could be related to the notch in the SiO disk observed at 7\,AU (Sect.~\ref{molecB}).
The polarimetric signature of plume \#2 shows that it contains dust, and the high degree of linear polarization ($p_L \approx 30\%$) is characteristic of a large scattering angle, probably around $50^\circ$.
It is unresolved spatially in the SPHERE/ZIMPOL images, implying that it is less than $\approx 1$\,AU across, but its extension perpendicularly to the plane of the disk reaches at least 15\,AU.
Assuming that source B is at the origin of the generation of plume \#2, this very strong focusing is difficult to explain if B is simply a clump of gas and dust.
As a side note, we observe a signature of plume \#2 also in the $^{29}$SiO emission map presented in Fig.~\ref{29SiO-image} (left panel), as an elongated north-south emission emerging from the disk at a radius of 2\,AU.
The accretion of the AGB star wind and dust on a compact body appears as a plausible explanation for the origin of the plume and its coincidence with L$_2$\,Pup~B.
Jets are a typical signature of accretion in vastly different objects, from young stars \citepads{2007prpl.conf..231R, 2016arXiv160708645L} to active galactic nuclei \citepads{Doeleman355}.
Our proposed geometry of the plume with respect to the disk and source B is presented in Fig.~\ref{l2pupsketch}.
We speculate that the plume is created by an accretion disk around source B, and launched perpendicularly to the dust disk plane.
Considering the expected size of the Roche lobe of L$_2$\,Pup~B, the radius of this accretion disk is probably smaller than 0.3\,AU, that is,~5\,mas in radius.
It is therefore unresolved in our ALMA observations in band 7, as observed in the subtracted PVD profiles in the CO lines (Fig.~\ref{COsubmaps}).
However, future observations with ALMA in band 9 or 10 could reach a sufficient angular resolution ($\approx 5$\,mas) to resolve this putative accretion disk around L$_2$\,Pup~B.

The loop is the dominant feature in the envelope of L$_2$\,Pup in the thermal infrared domain.
It extends radially up to the eastern edge of the dust disk and expands vertically to the north up to approximately 10\,AU above the disk plane as a streamer.
We propose that the development of the loop is linked to the presence of source B, and that it follows the geometry represented in Fig.~\ref{l2pupsketch}.
A possible mechanism for the formation of the dust is that it condenses in the shadow of source B, where the lower temperature would be more favorable to the survival of dust grains \citepads{2007MNRAS.376L...6M}.
A relatively dense accretion disk around L$_2$\,Pup~B could also favor the formation of dust.
The dust would then be blown away by the wind and radiation pressure of the AGB star, creating a planar spiral.
The three-dimensional development of the spiral into a streamer may be caused by the focusing of the wind by the flared disk along the polar axis, or by polar wind emitted by the disk itself.

\subsubsection{Relation with the disk and bipolar structure}

The overall morphology of the nebula of L$_2$\,Pup consists in an equatorial disk and bipolar cones \citepads{2015A&A...578A..77K}.
\citetads{2006MNRAS.370.2004N} showed that a low-mass companion in the envelope of a mass-losing star can lead to the formation of a disk and is likely to generate polar outflows.
\citetads{1999ApJ...523..357M} constructed three-dimensional hydrodynamical models of the dusty wind geometry in binary systems.
They obtained a range of envelope geometries from spherical to bipolar, including with internal spiral shock structures created by the orbital motion of the companion.
Specifically focused on L$_2$\,Pup, the simulations by \citetads{2016MNRAS.460.4182C} also indicate that an orbiting companion can result in the formation of an equatorial disk.
Possible traces of the interaction of the companion with the disk and bipolar cones can be observed in the visible images of L$_2$\,Pup as spirals and streamers (Fig.~\ref{ZIMPOL-image}; \citeads{2015A&A...578A..77K}).
This means that the general morphology of the envelope of L$_2$\,Pup is compatible with the expected signatures of a compact companion.

Owing to its near-Keplerian rotation and axial symmetry, the disk surrounding L$_2$\,Pup holds a considerable angular momentum.
The origin of this momentum cannot be explained by the current rotational velocity of the star, however, which is most likely very slow because of its strong inflation on the red giant branch \citepads{2009ApJ...695..679M}.
As discussed by \citetads{2016MNRAS.460.4182C}, a planetary mass or brown dwarf companion can inject angular momentum into the circumstellar disk and shape the wind of the star.
The compact body thus provides a simple explanation to the angular momentum of the disk, while a fluffy aggregate of gas and dust does not.

\subsection{Scenarios for the formation and evolution of L$_2$\,Pup~B}

L$_2$\,Pup~B could be an old planet that formed together with the star, or alternatively a second-generation body recently formed in the disk \citepads{2007ApJ...662..651I}.
Its uncertain mass is compatible with a low-mass rocky object that could have formed relatively quickly in the circumstellar disk, even though its estimated total dust mass is relatively low (\citeads{2015A&A...578A..77K} estimated a few $10^{-7}\,M_\odot$).
\citetads{2014MNRAS.437.3288B,2013MNRAS.431.3025B} reported debris disks around subgiants, which may provide seeds for the formation of second-generation planetary bodies.
A circumbinary dust disk and probably also planets orbiting the post-common envelope binary \object{NN Ser} was reported by \citetads{2016MNRAS.459.4518H} (see also \citeads{2016RSOS....350571V}).
This strengthens the credibility of the scenario of second-generation planet formation. 

The dynamical evolution of L$_2$\,Pup~B is essentially conditioned by its mass, the mass of the central star, the stellar mass-loss rate and tidal interactions.
%
An overview of the two-body mass-loss problem is presented in \citetads{2011MNRAS.417.2104V}.
If the mass of L$_2$\,Pup~B is in the gaseous giant or brown dwarf regime, the large convective envelope of the AGB star induces strong tidal forces.
\citetads{2012ApJ...761..121M} predicted that these forces will pull the planet inside the envelope of the AGB star if their initial orbital radius is shorter than 3\,AU.
This has also been the conclusion of \citetads{2007ApJ...661.1192V}.
The tidal stability of any exoplanetary system strongly depends on the tidal quality factor $Q$ \citepads{2014MNRAS.445.4137M}, which is poorly calibrated for main-sequence-star -- Jupiter-mass systems, and essentially unconstrained for evolved systems.
With a current orbital radius of 2.4\,AU, L$_2$\,Pup~B may currently be migrating toward the star.
\citetads{2010MNRAS.408..631N} predicted that once engulfed, Jupiter-mass companions will be destroyed during the common-envelope phase and therefore will not remain in orbit around the final white dwarf.
Lower mass planets are more likely to survive as their orbital radius will expand as the star loses mass.
Accretion onto the companion or evaporation \citepads{2009ApJ...705L..81V}, anisotropic mass loss or jets \citepads{2013MNRAS.435.2416V} and viscous interaction with the disk \citepads{1996Natur.380..606L} will also influence the companion's orbital evolution.
While tidal factors should circularize the orbit relatively quickly, eccentricity pumping of the companion's orbit through interactions with the disk are likely to have a destabilizing influence.
The signature of such interactions may be apparent both in the orbital parameters of L$_2$\,Pup~B (in particular eccentricity and inclination with respect to the disk plane) and in the disk structure (warp).

Considering L$_2$\,Pup as an analog of the future Sun, the present properties of L$_2$\,Pup indicate that Mercury and Venus will very likely be engulfed in the Sun's envelope as they orbit within or very close to the present radius of the AGB star (0.6\,AU).
Assuming that the orbital energy of L$_2$\,Pup~B remained constant since its formation, its orbital radius while component A was on the main sequence (when $m_A \approx 1\,M_\odot$) was around 1.6\,AU, that is, comparable to the present orbit of Mars around the Sun.
The presence of L$_2$\,Pup~B is thus consistent with the minimum radius of 1.15\,AU inferred by \citetads{2008MNRAS.386..155S} for the survival of planets orbiting the Sun.
The orbital interactions between the planets will play a major role in determining their survival.
For instance, an inward migration of Jupiter would naturally have a major impact on the inner planets.
The fate of Earth also strongly depends on the detail of how the RGB and AGB phases progress, particularly with regard to asphericities in the stellar wind and the timing and efficiency of RGB/AGB mass loss.
Different models prescribe different outcomes Earth, and L$_2$\,Pup could provide important indications to distinguish between its engulfment in the central star and its survival as a white dwarf planet.

\section{Conclusion}

From the Keplerian rotation of $^{29}$SiO molecular gas observed with ALMA, we determined that the mass of the central object of L$_2$\,Pup is $m_A = \mass \pm\masserrstat \pm \masserrparallax\,M_\odot$.
The error budget ($\pm \masserrpercent$) is fully dominated by the uncertainty of the \emph{Hipparcos} parallax, which will soon be improved by \emph{Gaia}.
This accurate mass combined with the other observed properties of the AGB star (pulsation period, radius luminosity, etc.) allowed us to conclude from evolutionary models that the mass of L$_2$\,Pup~A when it was on the main sequence is close to solar, and that its current age is approximately 10\,Gyr.
This age is consistent with the Galactic space velocity of the star, which indicates that it is probably a member of the thick-disk population.

We also identified a secondary source of continuum and molecular emission, located at a projected radius of $\approx 2$\,AU.
This position corresponds to the location of the companion L$_2$\,Pup~B reported by \citetads{2015A&A...578A..77K}.
From its estimated mass of $\massBjup \pm \massBjuperr\,M_\mathrm{Jup}$ and observed emission, we argue that source B is either a planet or a low-mass brown dwarf accreting the wind of the AGB star.
While it could formally be a dense clump of dust and molecules, its persistence over one year, its coincidence with remarkable nebular features and the overall morphology of the envelope of L$_2$\,Pup all favor the hypothesis of a compact body.
We emphasize, however, that the properties of source B are still uncertain as we do not have a firm lower limit on its mass.
The hypothesis that it is a dense clump of dust and gas cannot be formally excluded.
Very high angular resolution observations with ALMA at short wavelengths may resolve the putative accretion disk surrounding L$_2$\,Pup~B, and allow a more precise measurement of its mass.
A schematic view of the configuration we propose for the environment of L$_2$\,Pup is presented in Fig.~\ref{l2pupsketch}.

From its observed properties, L$_2$\,Pup and its companion emerge as a plausible analog of the solar system at an age of approximately 10\,Gyr.
It provides a view on the complex interactions occurring between a solar-type star entering the planetary nebula phase and its planetary system.
The companion could also play an important role in the shaping of the bipolar envelope of L$_2$\,Pup and subsequently of the planetary nebula \citepads{2006MNRAS.370.2004N}.
Future observations of L$_2$\,Pup with ALMA's highest angular resolutions and the E-ELT, for instance, will provide valuable constraints for the modeling of these interactions (see e.g.~\citeads{2016MNRAS.458..832S, 2016MNRAS.460.4182C, 2015A&A...579A.118H, 2014ASPC..487..401W}).

\begin{acknowledgements}
This article makes use of the following ALMA data: ADS/JAO.ALMA\#2015.1.00141.S . ALMA is a partnership of ESO (representing its member states), NSF (USA) and NINS (Japan), together with NRC (Canada), NSC and ASIAA (Taiwan), and KASI (Republic of Korea), in cooperation with the Republic of Chile. The Joint ALMA Observatory is operated by ESO, AUI/NRAO and NAOJ.
WH acknowledges support from the Fonds voor Wetenschappelijk Onderzoek Vlaanderen (FWO).
LD acknowledges support from the ERC consolidator grant 646758 AEROSOL and the FWO Research Project grant G024112N.
IM acknowledges support from the UK Science and Technology Research Council, under grant number ST/L000768/1
We acknowledge financial support from the ``Programme National de Physique Stellaire" (PNPS) of CNRS/INSU, France.
This research received the support of PHASE, the high angular resolution partnership between ONERA, Observatoire de Paris, CNRS and University Denis Diderot Paris 7.
We acknowledge with thanks the variable star observations from the AAVSO International Database contributed by observers worldwide and used in this research.
This research made use of Astropy\footnote{Available at \url{http://www.astropy.org/}}, a community-developed core Python package for Astronomy \citepads{2013A&A...558A..33A}.
We used the SIMBAD and VIZIER databases at the CDS, Strasbourg (France), and NASA's Astrophysics Data System Bibliographic Services.
\end{acknowledgements}

\bibliographystyle{aa} 
\bibliography{biblioL2Pup2016}

\begin{appendix}

\section{$^{12}$CO$(\varv=0,J=3-2)$ line\label{12COline}}

\begin{figure*}[]
        \centering
        \includegraphics[width=8cm]{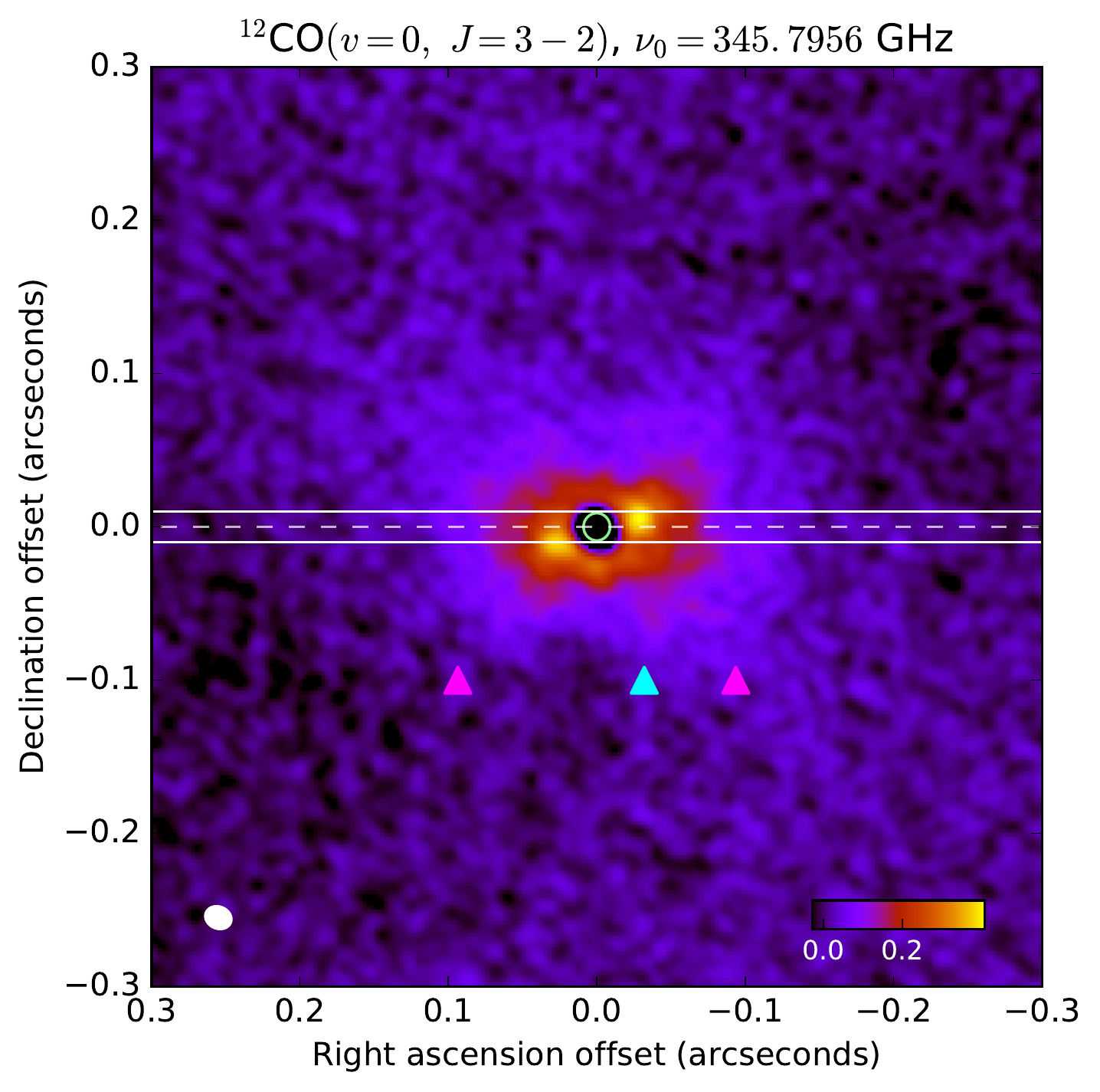}
        \includegraphics[width=8cm]{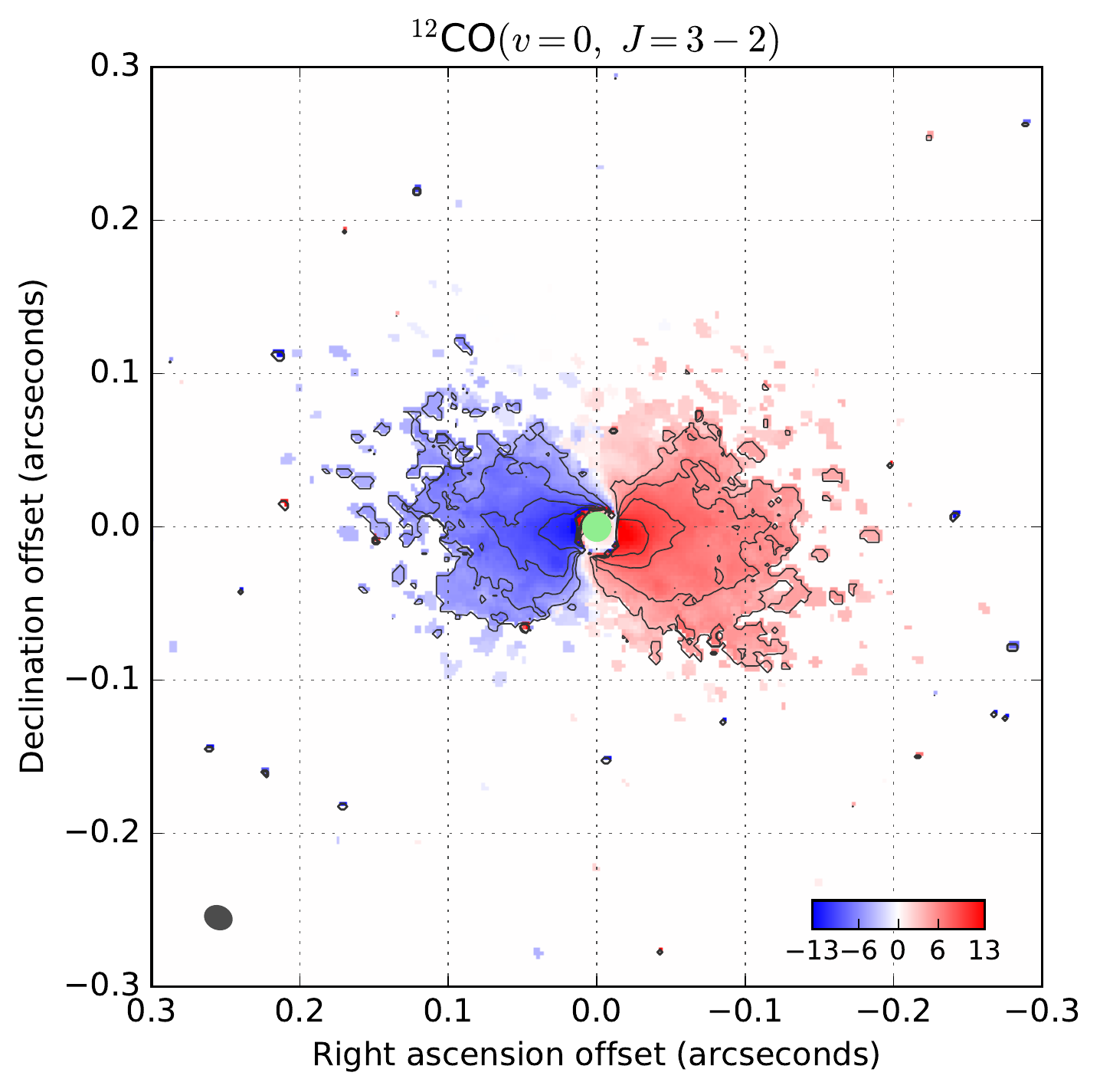}
        \includegraphics[width=16cm]{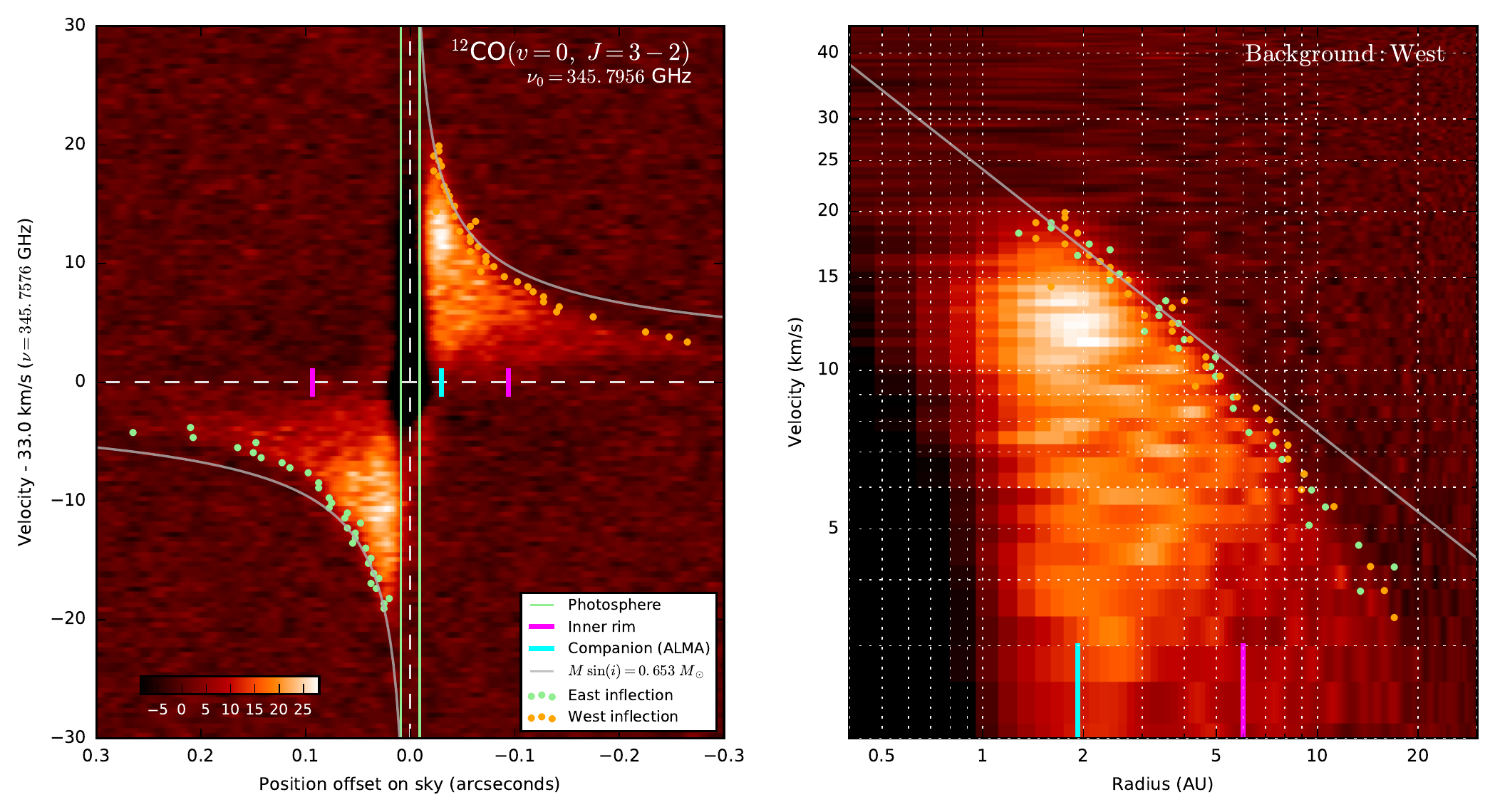}
        \caption{{\it Top left panel:} Map of the emission from L$_2$\,Pup in the $^{12}$CO$(\varv=0,J=3-2)$ line.
        {\it Top right panel:} First moment of velocity map (color scale in km\ s$^{-1}$). The contours are drawn between 4 and 10\,km\,s$^{-1}$ with a 2\,km\,s$^{-1}$ step, and the size of the photosphere is shown with a light green disk.
        {\it Bottom panels:} Position-velocity diagram of the $^{12}$CO$(\varv=0,J=3-2)$ emission line.
        The Keplerian velocity profile corresponding to a central mass of $M \sin(i) = \masssini\ M_\odot$ (determined from the $^{29}$SiO line fit) is shown with gray curves in the two panels.
        \label{12CO-line}}
\end{figure*}

The emission of L$_2$\,Pup in the $^{12}$CO line is presented in Fig.~\ref{12CO-line} (top panel).
The total peak intensity is 0.35 to 0.40\,Jy\ beam$^{-1}$\ km\ s$^{-1}$ and the maximum emission is located at a radius of approximately 2\,AU. The degree of asymmetry is small, although the position angles of the emission peaks are slightly tilted with respect the position angle of the plane of the disk.
We detect $^{12}$CO$(v=0,J=3-2)$ emission between -20 and +20\,km\,s$^{-1}$ relative to the mean velocity of L$_2$\,Pup ($+33.0$\,km\,s$^{-1}$).
Within a 0.63\,arcsecond aperture, we measure an emission peak of 5.4\,Jy, centered close to the stellar velocity.
Single-dish observations by \citetads{1999A&AS..138..299K} showed spectral peaks that are several times higher than ALMA in the same
transition of CO but only close to the stellar velocity.
This suggests that in addition to the high-velocity CO detected by ALMA close to the star, there exists an extended, slowly expanding, diffuse envelope that is resolved out by the long baselines of the interferometer.

\section{$^{13}$CO$(\varv=0,J=3-2)$ line\label{13COline}}

\begin{figure*}[]
        \centering
        \includegraphics[width=8cm]{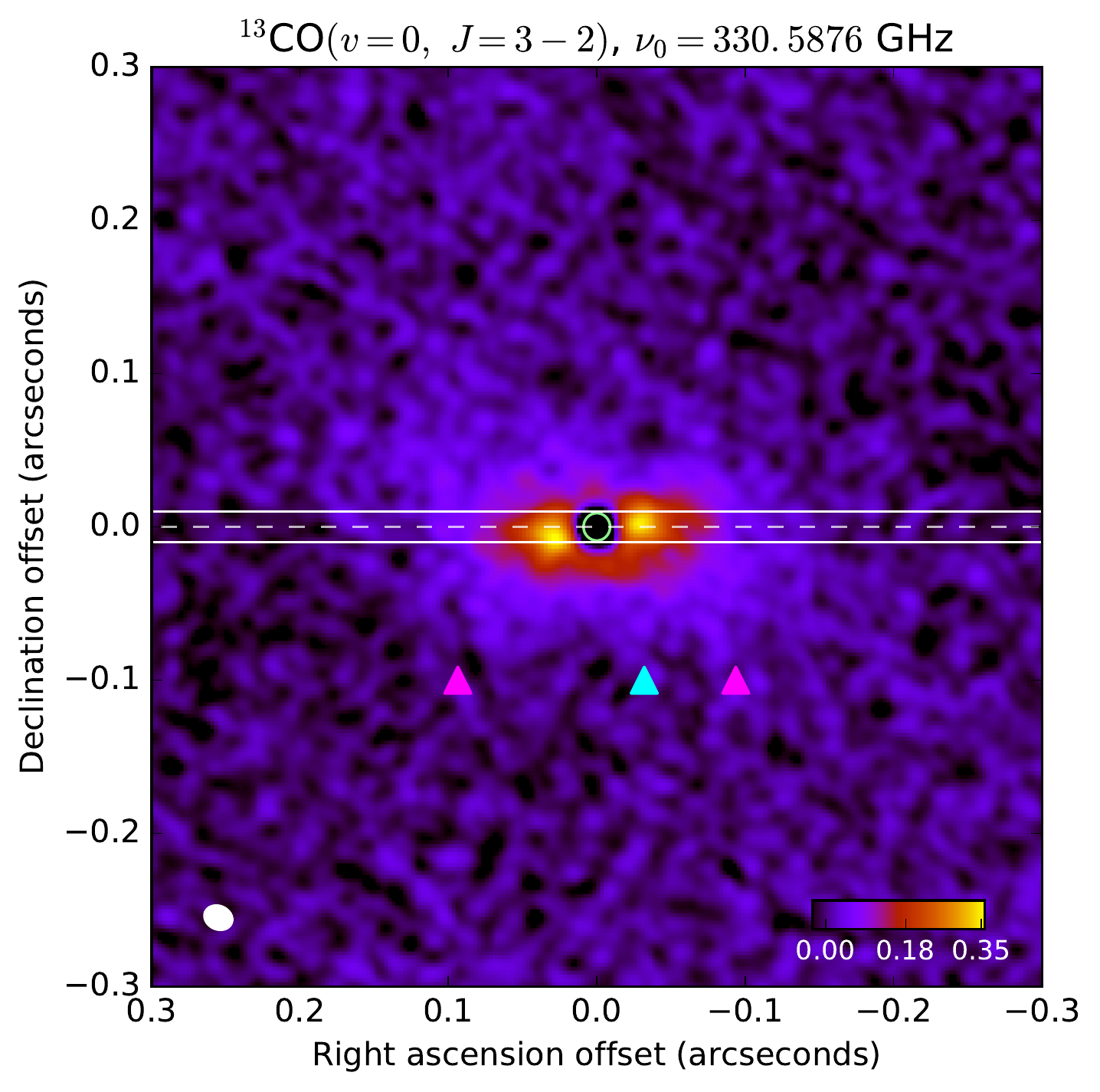}
        \includegraphics[width=8cm]{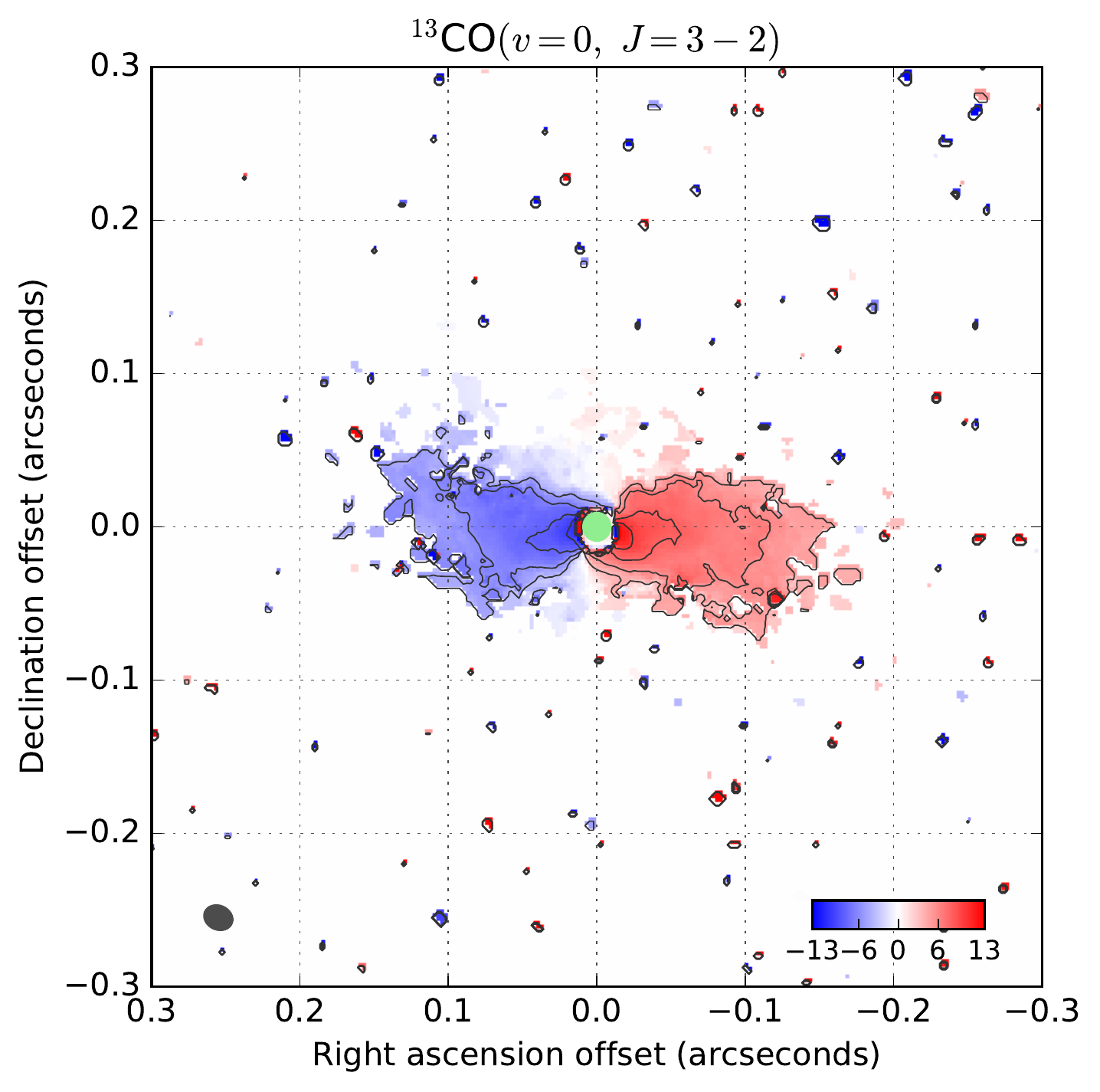}
        \includegraphics[width=16cm]{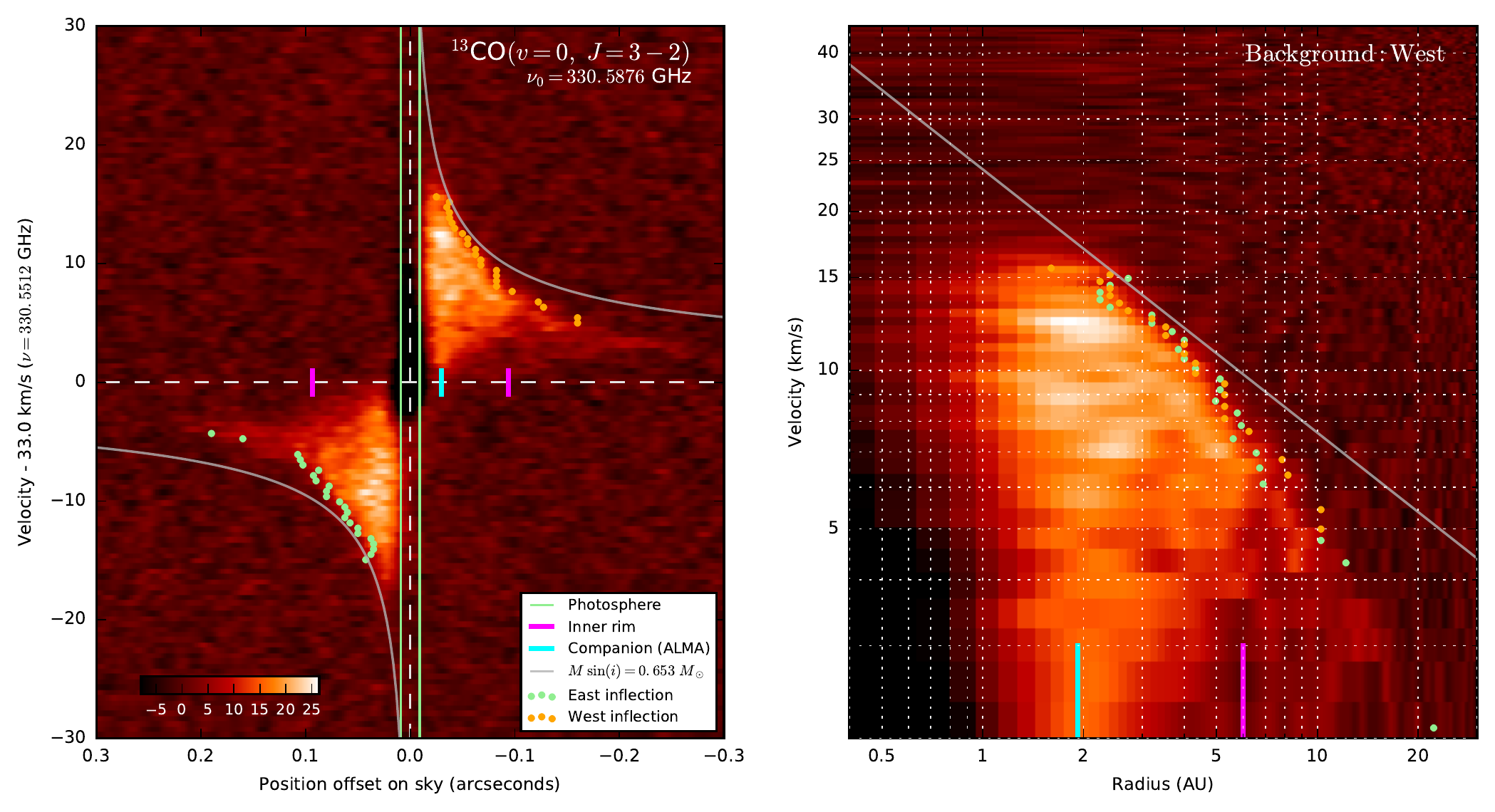}
        \caption{{\it Top left panel:} Map of the emission from L$_2$\,Pup in the $^{13}$CO$(\varv=0,J=3-2)$ line.
        {\it Top right panel:} First moment of velocity map (color scale in km\ s$^{-1}$).
        {\it Bottom panels:} Position-velocity diagrams.
        \label{13CO-line}}
\end{figure*}

The $^{13}$CO emission of L$_2$\,Pup (Fig.~\ref{13CO-line}, top panel) has a similar radial extension to the $^{12}$CO line but the $^{13}$CO isotopologue appears significantly more confined in the plane of the disk.
The total peak intensity is comparable to $^{12}$CO (0.35 to 0.40\,Jy\ beam$^{-1}$\ km\ s$^{-1}$) and the maximum emission is also located at a radius of approximately 2\,AU.
The position angles of the emission peaks is less tilted than $^{12}$CO with respect the east-west axis of the disk.

\section{SO$_2(\varv=0,34(3,31)-34(2,32))$ line\label{SO2line}}

\begin{figure*}[]
        \centering
        \includegraphics[width=8cm]{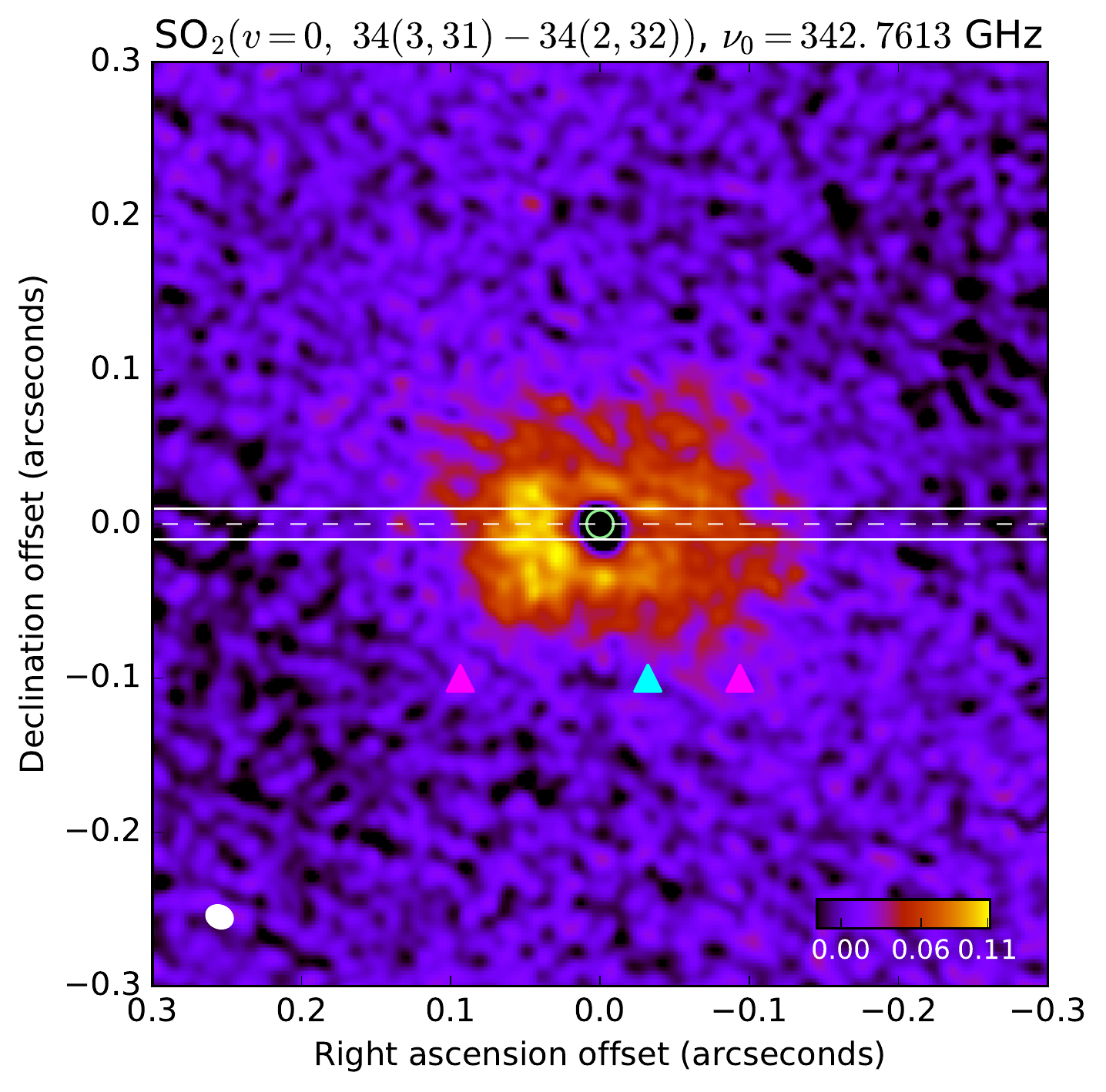}
        \includegraphics[width=8cm]{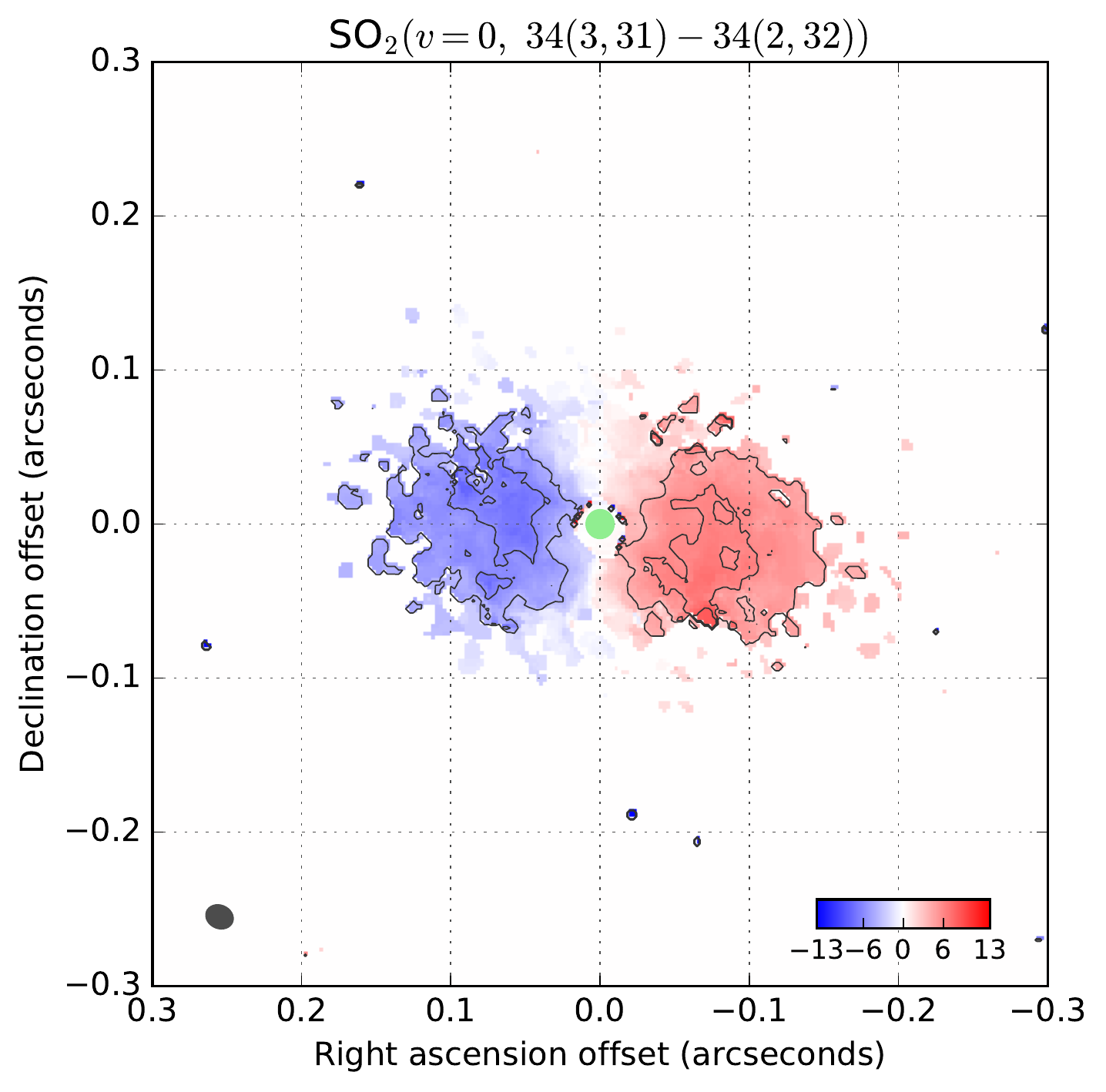}
        \includegraphics[width=16cm]{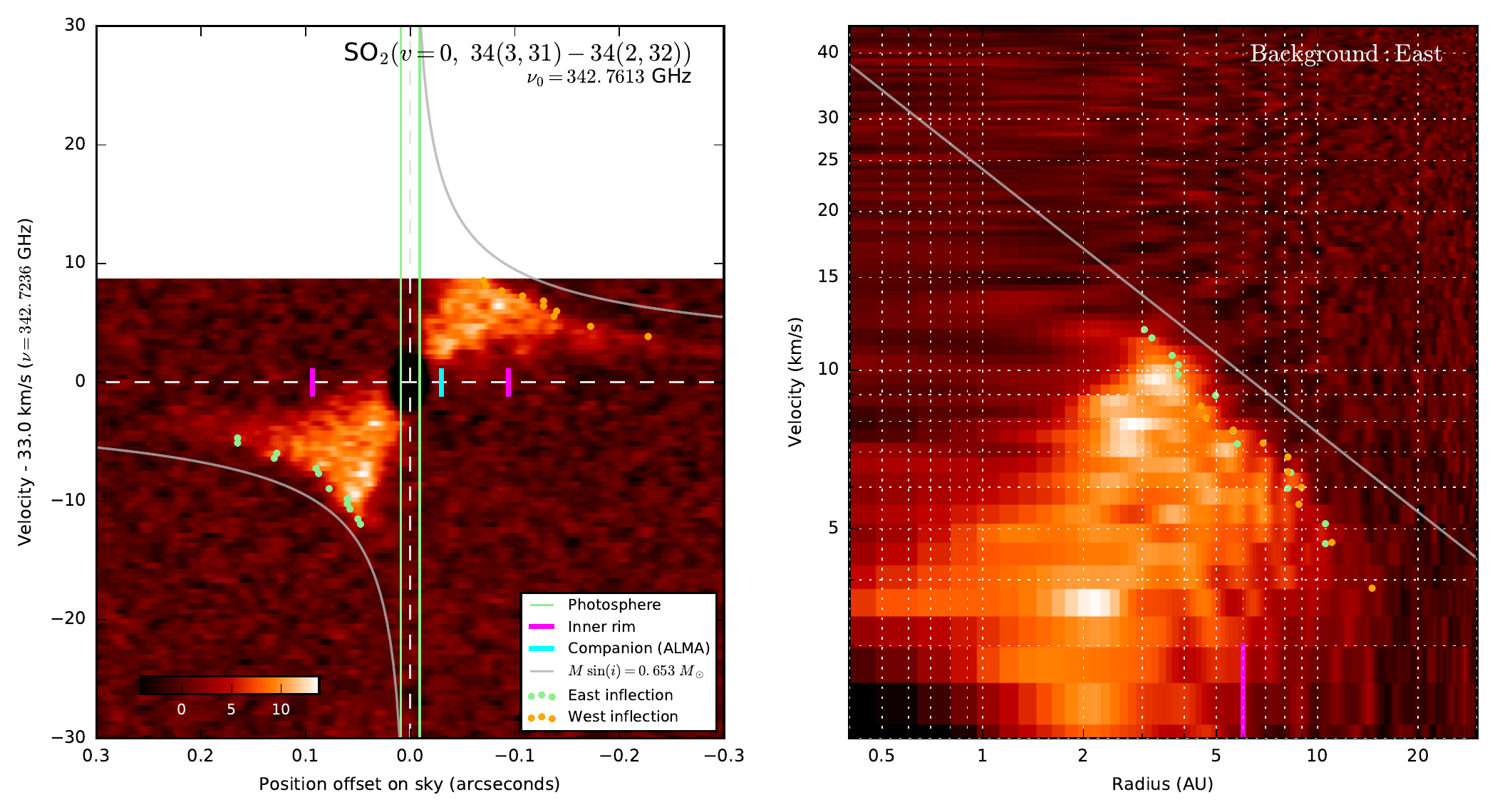}
        \caption{Emission map ({\it top left panel}), first moment of velocity map ({\it top right panel}) and position-velocity diagram ({\it bottom panels}) of the SO$_2(\varv=0,34(3,31)-34(2,32))$ emission line.
        \label{SO2-pvd}}
\end{figure*}

As shown in Fig.~\ref{SO2-pvd} (top panel), the sulfur dioxide emission is spread over a large vertical extension of approximately $\pm 0.1\arcsec$ with respect to the disk plane.
The PVD is incomplete as the line was located close to the edge of the spectral window.
In the flux image, the emission appears stronger on the east side of the disk, but this is an artifact due to the incompleteness of the coverage of the frequencies showing Doppler shifted emission.
The velocity profile is sub-Keplerian over the full extent of the detected emission, but the deviation is smaller as the radius decreases.
At a radius of 3\,AU, the deviation from the Keplerian velocity is approximately $\Delta v = -2$\,km\,s$^{-1}$.

\section{SO$\ 3\Sigma\ (\varv=0,8(8)-7(7))$ line\label{SOline}}

\begin{figure*}[]
        \centering
        \includegraphics[width=8cm]{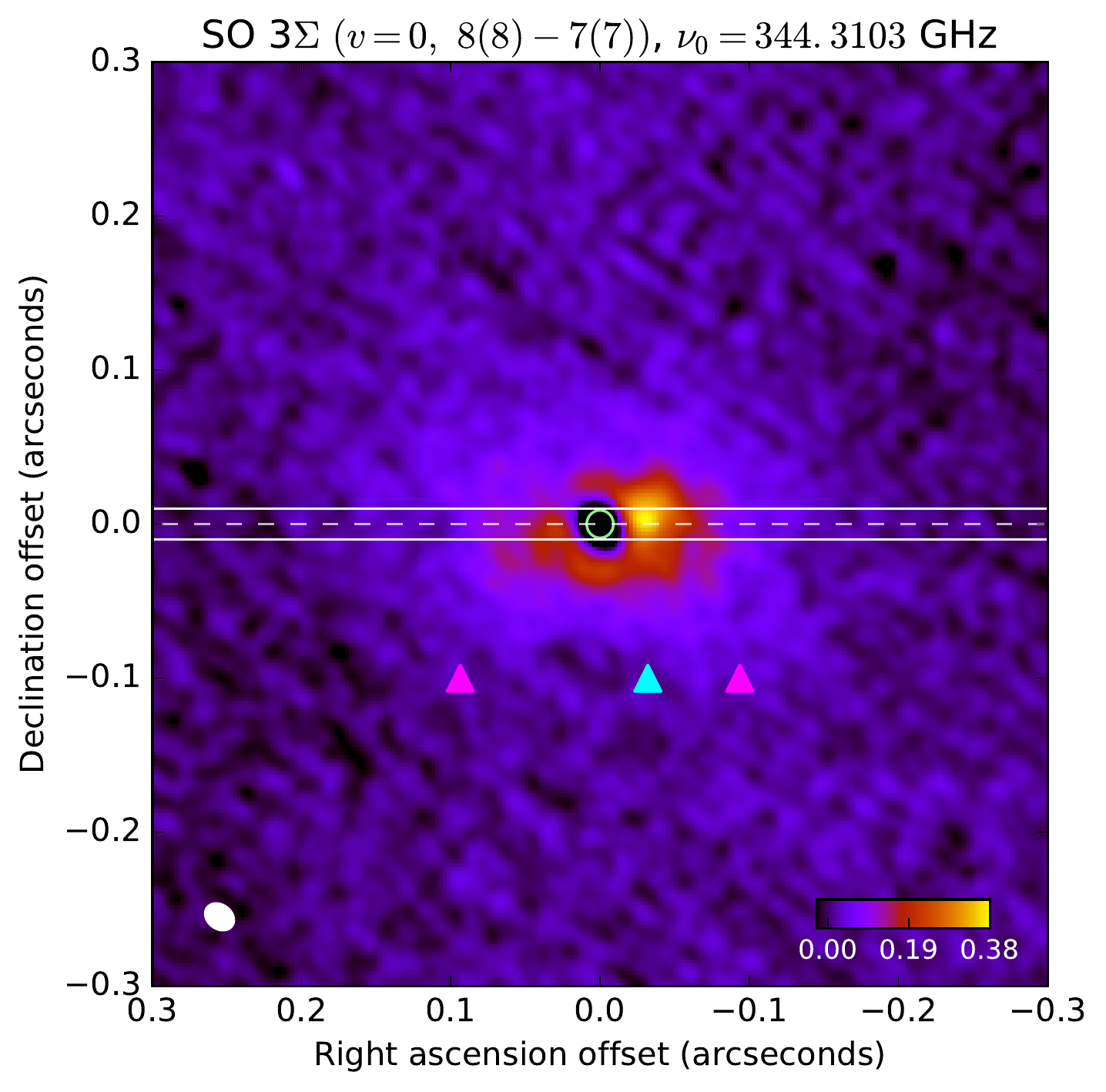}
        \includegraphics[width=8cm]{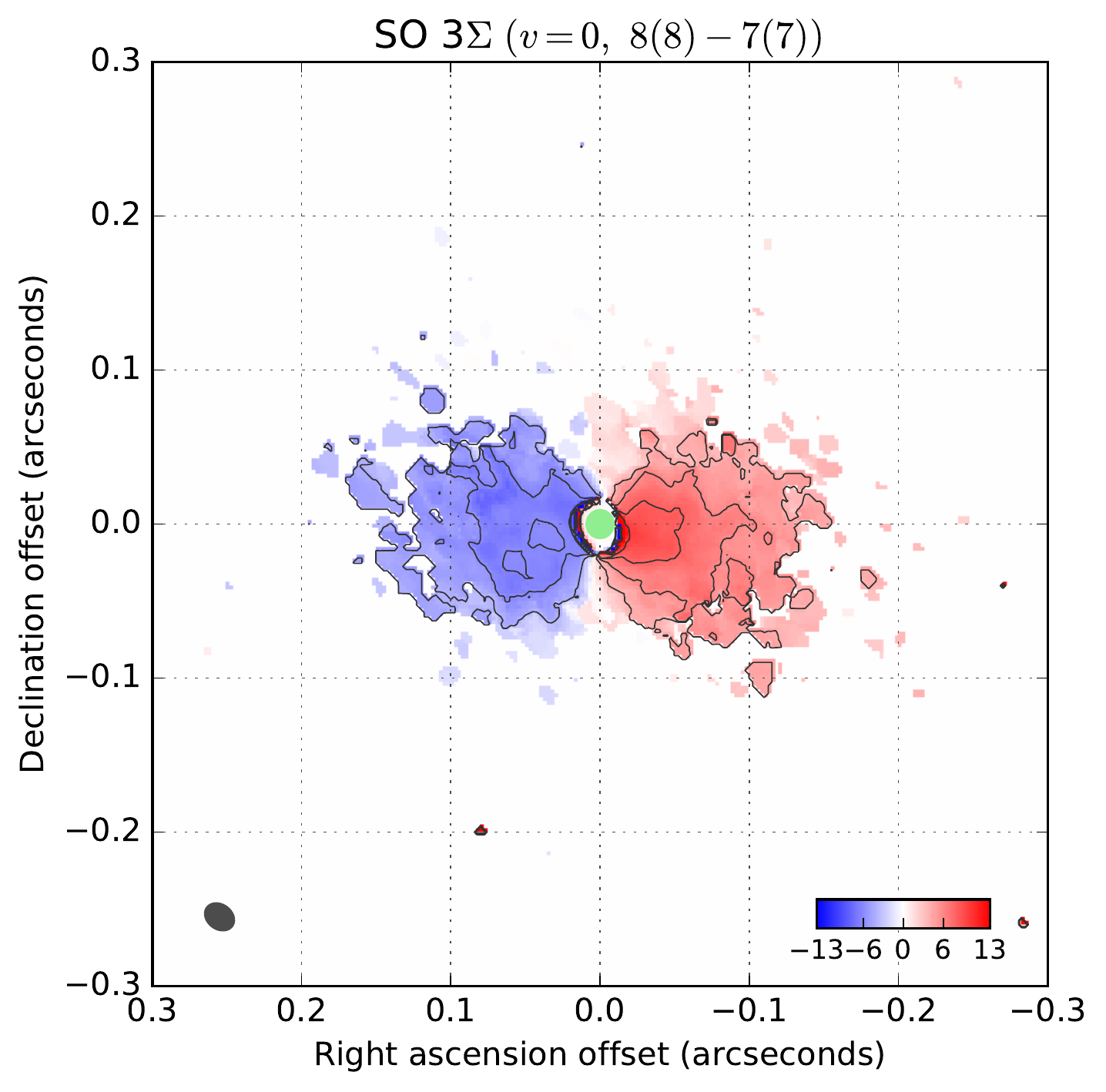}
        \includegraphics[width=16cm]{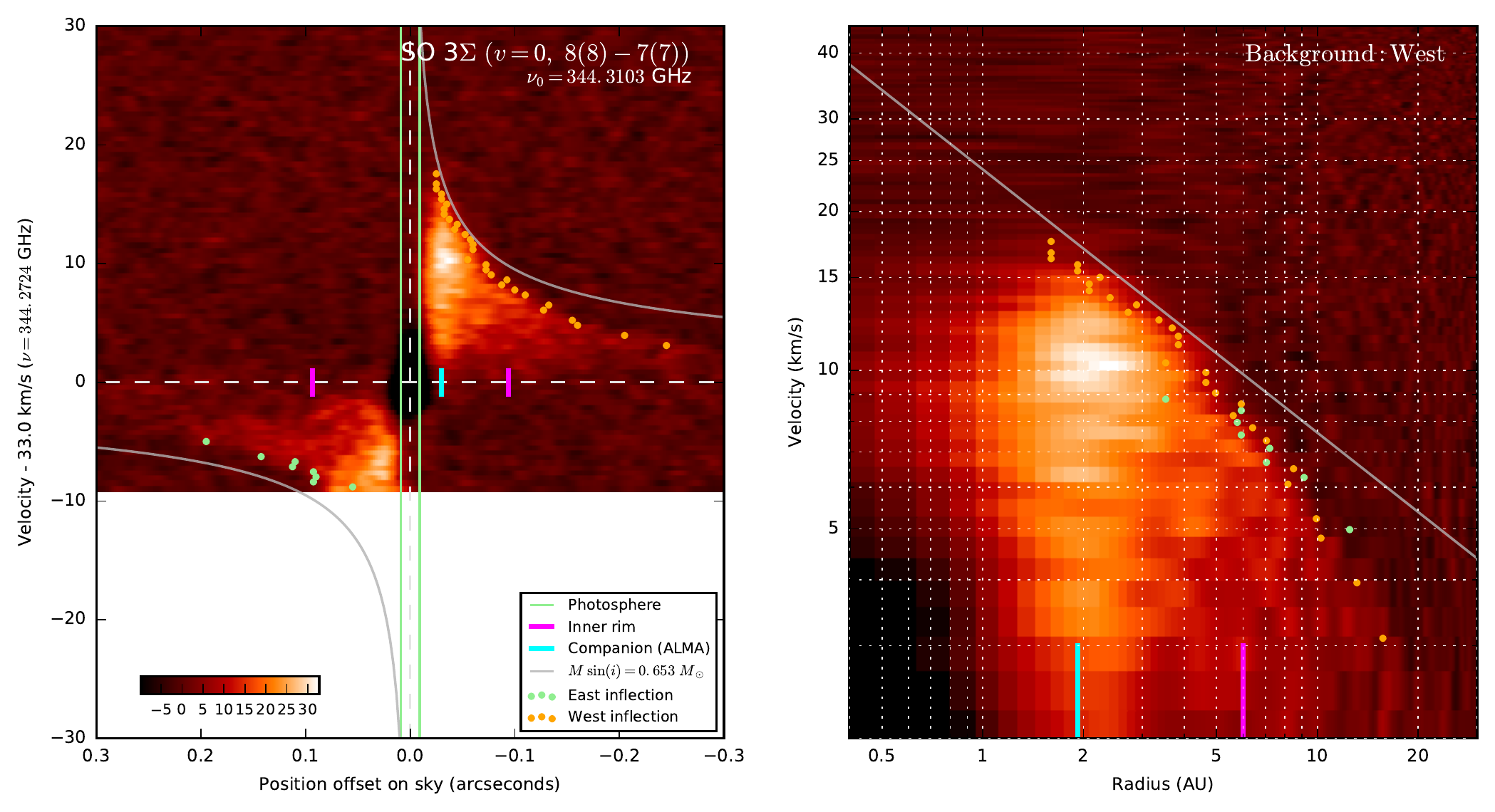}
        \caption{Emission map ({\it top left panel}), first moment of velocity map ({\it top right panel}) and position-velocity diagram ({\it bottom panels}) of the SO$\ 3\Sigma\ (\varv=0,8(8)-7(7))$ emission line.
        \label{SO-pvd}}
\end{figure*}

The SO$\ 3\Sigma\ (\varv=0,8(8)-7(7))$ emission map (Fig.~\ref{SO-pvd}) exhibits a strong emission reaching 0.18\,Jy\ beam$^{-1}$\ km\ s$^{-1}$ on the west side of the disk, close to the position of the L$_2$\,Pup~B source (Sect.~\ref{companion}).
As for the SO$_2$ line (Appendix~\ref{SO2line}), the PVD is incomplete as the line was located close to the edge of the spectral window, and this explains the apparent east-west asymmetry of the flux distribution.
The western part of the PVD represented in the bottom right panel of Fig.~\ref{SO-pvd} shows a rotation velocity close to Keplerian between 2 and 4\,AU from the central star.

\section{SiS$(\varv=1,J=19-18)$ line\label{SiSline}}

\begin{figure*}[]
        \centering
        \includegraphics[width=8cm]{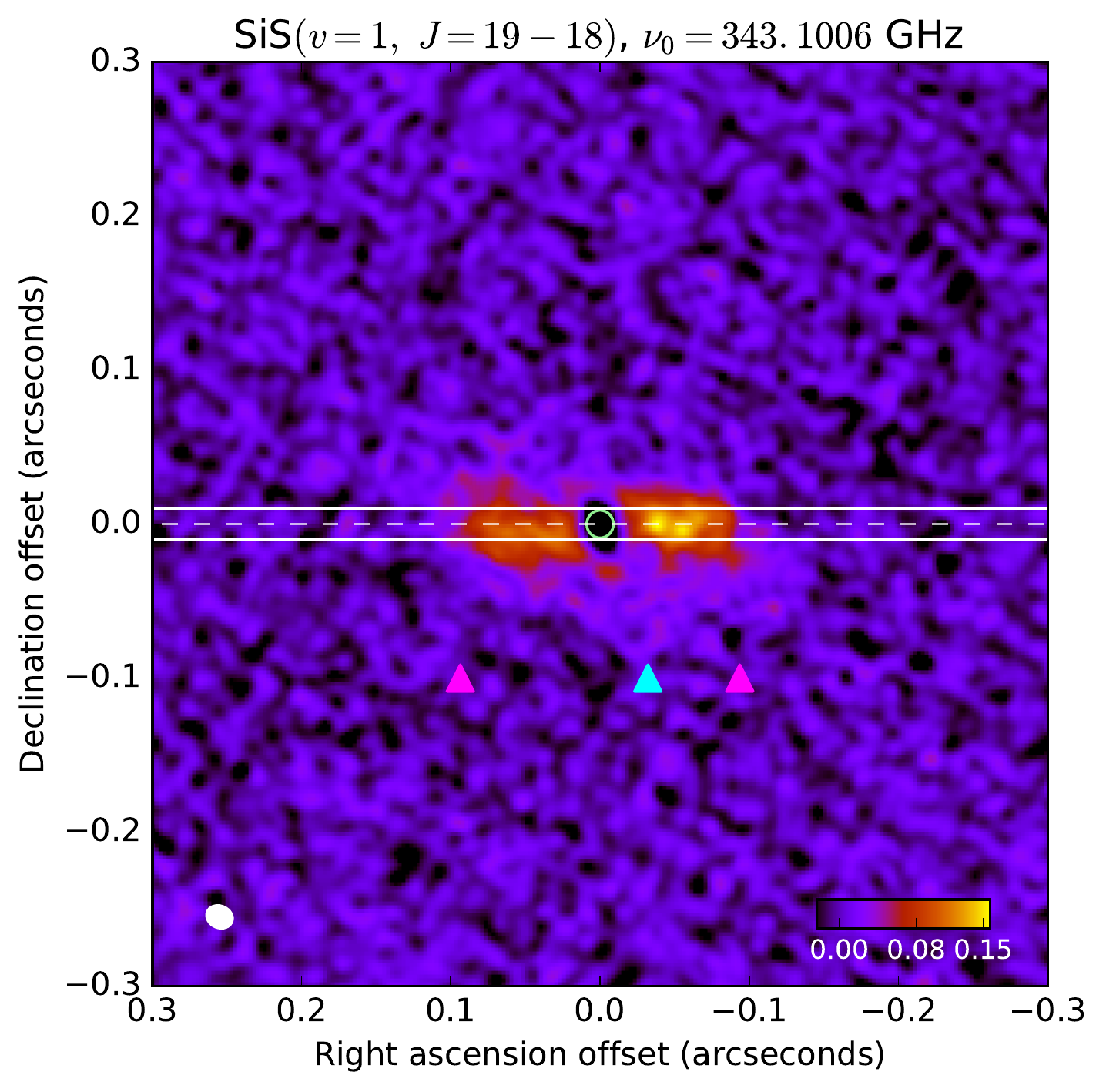}
        \includegraphics[width=8cm]{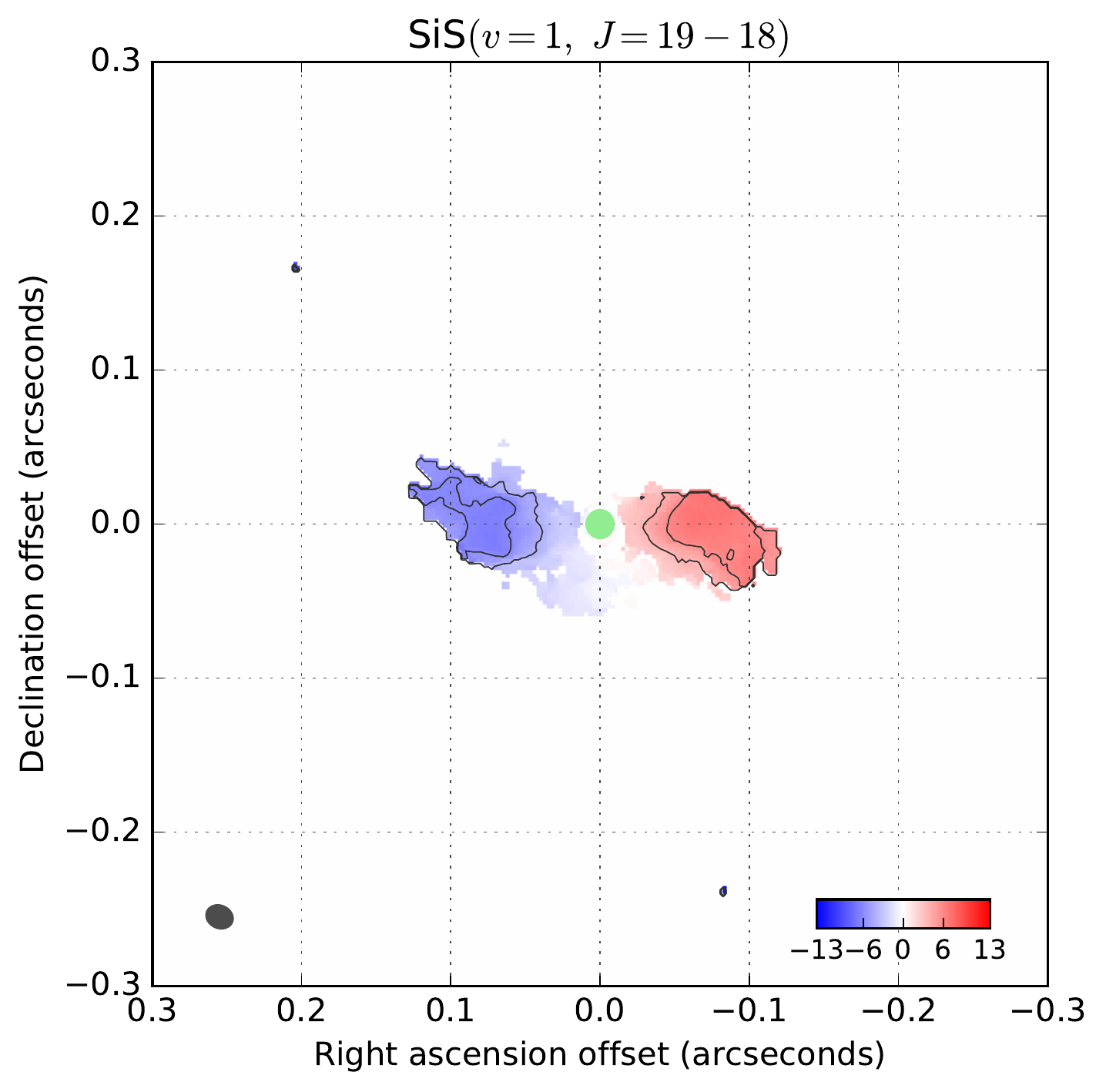}
        \includegraphics[width=16cm]{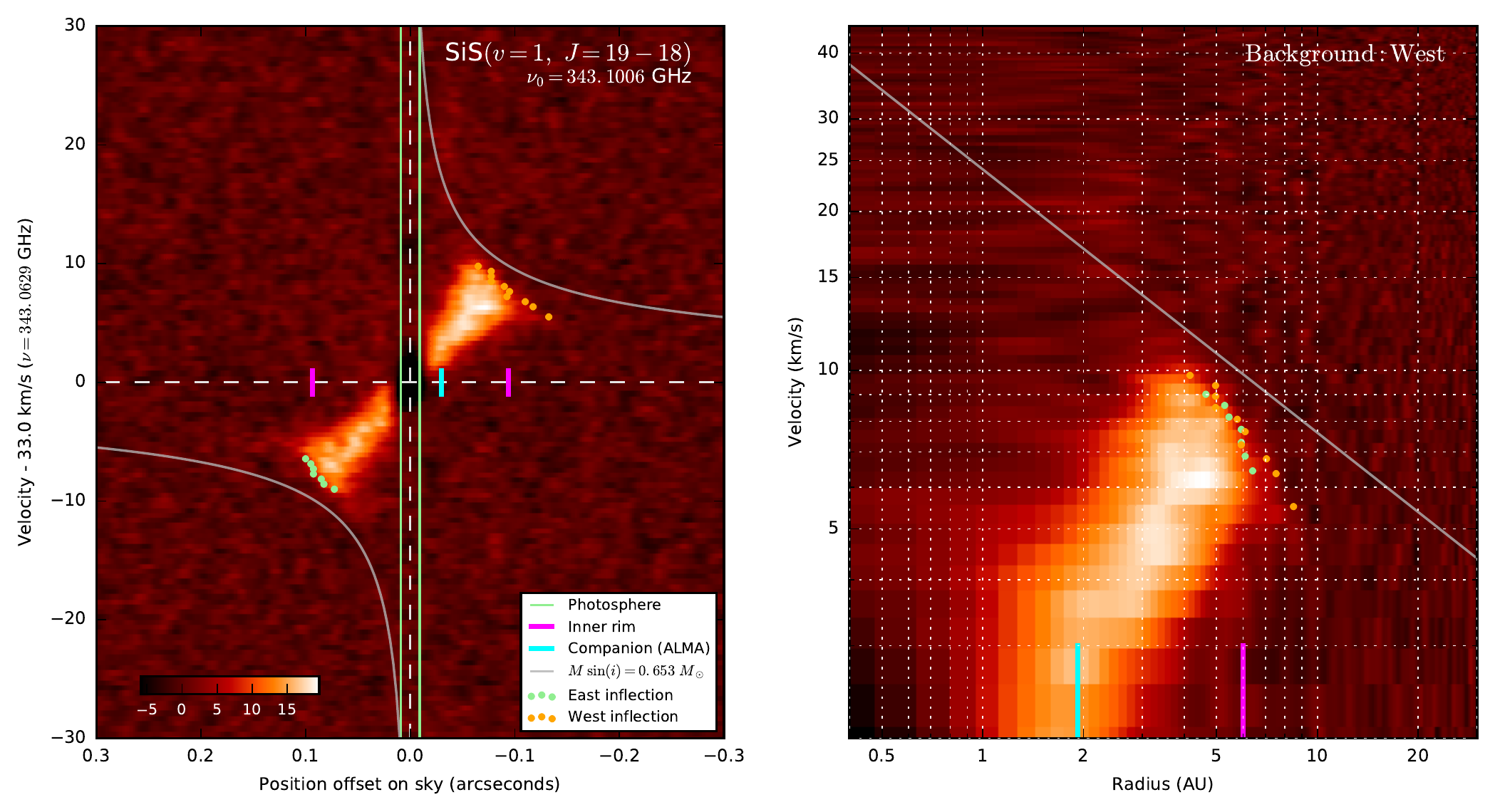}
        \caption{Emission map ({\it top left panel}), first moment of velocity map ({\it top right panel}) and position velocity diagram ({\it bottom panels}) of the SiS$(\varv=1,J=19-18)$ emission line.
        \label{SiS-line}}
\end{figure*}

The emission map in the SiS$(\varv=1,J=19-18)$ (Fig.~\ref{SiS-line}) shows that this line is narrowly contained within the plane of the dust disk and has a very restricted vertical extension.
The PVD indicates that the emission is confined to a thin ring located around the inner rim of the dust disk (6\,AU), and that its radial extension is mostly limited to between 4 and 7\,AU.
No significant emission is detected beyond a radius of approximately 10\,AU.
The PVD shows that the rotational velocity is always sub-Keplerian, with an increasing deviation from the $M \sin(i) =\masssini\,M_\odot$ velocity profile with increasing radius.
The emission is asymmetric between the east and west parts of the disk, with a significantly stronger emission on the west side.

\end{appendix}

\end{document}